\documentclass[11pt]{article}
\usepackage{amsmath,amssymb,amsthm}
\usepackage{mathrsfs,bm}
\usepackage{graphicx,epstopdf,caption,ulem}
\graphicspath{{figures/}}  

\usepackage{float,subcaption,setspace,booktabs,multirow,supertabular,lscape,threeparttable,diagbox}
\usepackage{colortbl}  
\usepackage{multirow}  
\usepackage[margin=1.0in]{geometry}  
\usepackage{setspace}

\usepackage[noblocks]{authblk}  
\usepackage{natbib}  
\usepackage{algorithm}
\usepackage{algorithmicx}
\usepackage{algpseudocode}
\usepackage[figuresright]{rotating}

\makeatletter
\def\app@number#1{ \setcounter{#1}{0}%
\@addtoreset{#1}{section}%
\@namedef{the#1}{\thesection.\arabic{#1}}}
\def\appendix{\@ifstar{\appendix@star}{\appendix@nostar}}
\def\appendix@nostar{%
\def\lb@section{ \appendixname \ \thesection.\half@em}
\def\lb@empty@section{\appendixname\ \thesection}
\setcounter{section}{0}\def\thesection{Appendix \Alph{section}}%
\app@number{equation}}
\def\appendix@star{%
\def\lb@section{\appendixname}\let\lb@empty@section\lb@section
\setcounter{section}{0}\def\thesection{Appendix \Alph{section}}%
\app@number{equation}\app@number{figure}}
\makeatother

\begin{document}
\title{Robust Online Detection in Serially Correlated Directed Network}
\author[a]{Miaomiao Yu}
\author[b]{Yuhao Zhou*}
\author[c]{Fugee Tsung}
\affil[a]{\footnotesize{Key Laboratory of Advanced Theory and Application in Statistics and Data Science, MOE, and Academy of Statistics and Interdisciplinary Sciences, East China Normal University, Shanghai, China}}
\affil[b]{\footnotesize{Department of Statistics and Operations Research, University of North Carolina at Chapel Hill, Chapel Hill, NC 27599, USA}}
\affil[c]{\footnotesize{Department of Industrial Engineering and Decision Analytics, Hong Kong University of Science and Technology, Kowloon, Hong Kong.}}
\date{}
\maketitle
{\bf Abstract:}
{As the complexity of production processes increases, the diversity of data types drives the development of network monitoring technology. This paper mainly focuses on an online algorithm to detect serially correlated directed networks robustly and sensitively. First, we consider a transition probability matrix to resolve the double correlation of primary data. Further, since the sum of each row of the transition probability matrix is one, it standardizes the data, facilitating subsequent modeling. Then we extend the spring length based method to the multivariate case and propose an adaptive cumulative sum (CUSUM) control chart on the strength of a weighted statistic to monitor directed networks. This novel approach assumes only that the process observation is associated with nearby points without any parametric time series model, which is in line with reality. Simulation results and a real example from metro transportation demonstrate the superiority of our design.}

\section{Introduction}
Statistical process control (SPC) is one of the most widely used tools to monitor the quality of manufactured products and is encountered in almost every field, including company operation analysis \citep{Han2020}, semiconductor manufacturing \citep{Xiang2021} and human genetics \citep{Zou2020}. In the era of big data, it faces tremendous challenges, especially in the sparse strategies for monitoring networks and/or graphics data that are prevalent in diverse scenarios \citep{Crane2018, Asikainen2020, Zhang2021}. In this paper, we focus on the problem of online monitoring a serially correlated directed network (e.g. transport system), in which correlation undermines the assumption of conventional monitoring techniques and leads to unreliable performance \citep{Qiu2021}.

\begin{figure}
  \centering
  \includegraphics[scale=0.5]{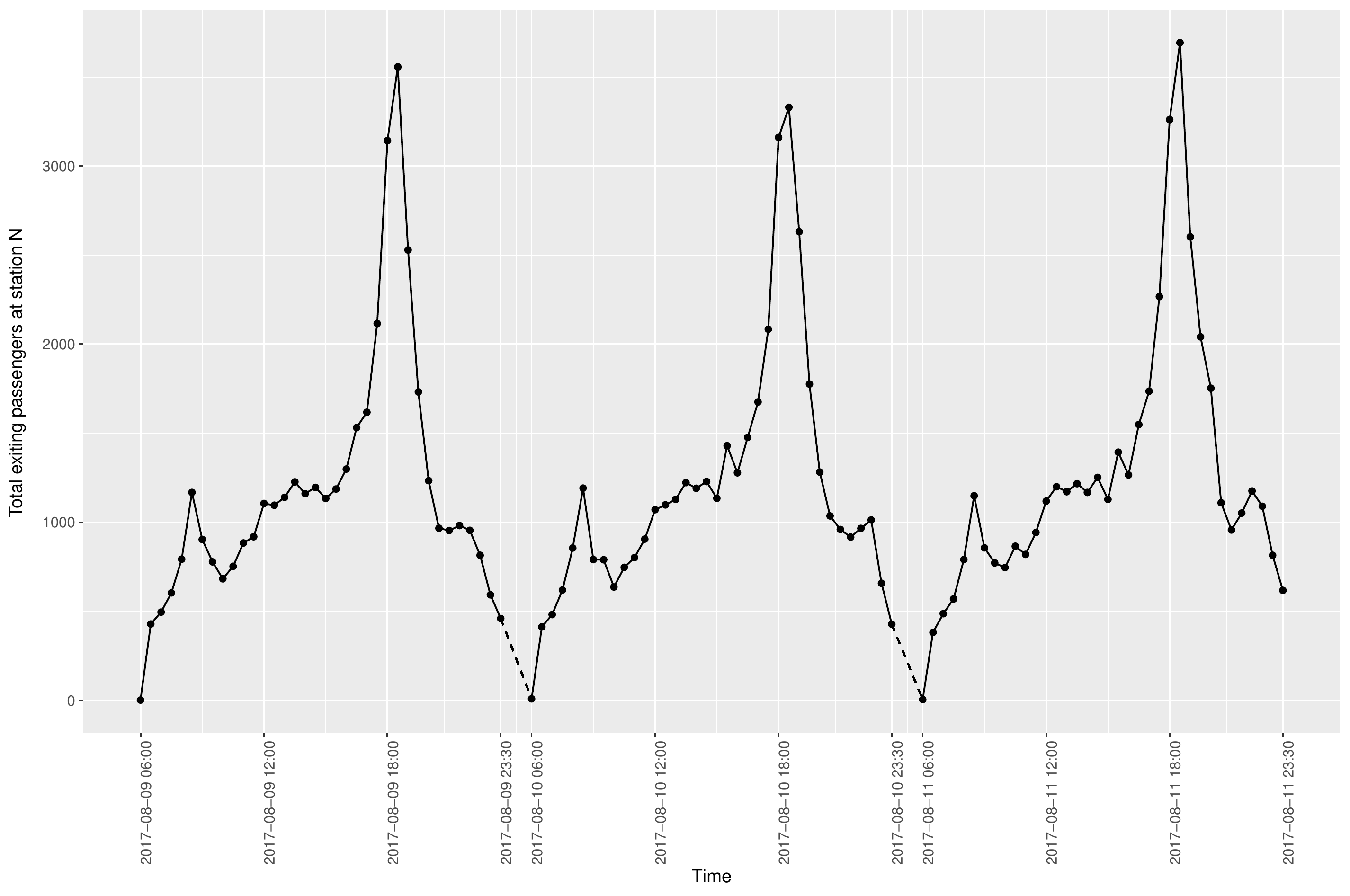}
  \caption{Traffic load at a certain station at half hour intervals.}\label{fig:number}
\end{figure}

\begin{figure}
  \centering
  \includegraphics[scale=0.5]{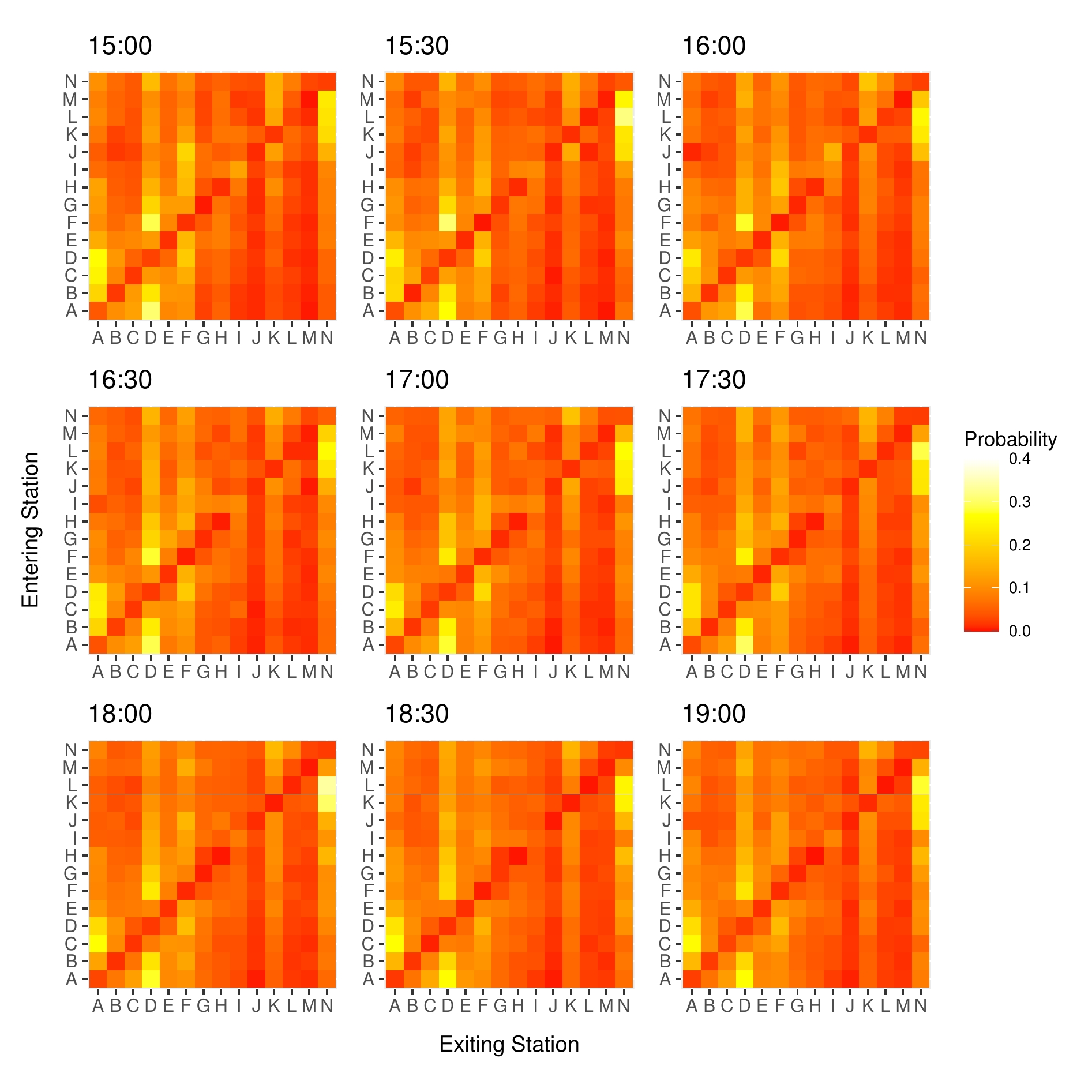}
  \caption{Heat map of some transition probability matrices in a particular line on August 10, 2017.}\label{fig:probability}
\end{figure}

The motivation of our paper comes from the real example of monitoring metro traffic. The urban metro transportation system, with its characteristics of reliability, high capacity and environmental friendliness, has become a popular means of transport that relieves traffic pressure in modern big cities. It is well known that the operations of metro lines are naturally unstable because of uncertain disturbances, such as system abnormalities, inadequate driver/passenger action, and so on \citep{Lin2010, Li2016}. Any deviation with respect to the nominal schedule of a given train will be amplified with time,  consequently disturbing the operation of other trains\citep{Campion1985}. Thus, it is important to check whether a particular line is in control (IC) and the negative results may remind operators to adjust schedules in time to avoid traffic jams and loss of property. In this case, there are two major problems. On the one hand, the metro traffic problem is a directed network that can be converted into a square matrix and few methods to detect the matrix are currently available. On the other hand, high-frequency traffic data does not satisfy the assumption of independence, which leads to decreasing algorithmic efficiency. What is worse, the existence of double correlation in this case challenges the stability of the algorithm even more greatly. More specifically, the serial correlation of the number of people transferred from station A to station B is caused by the autocorrelations of traffic load and transition probability as shown in Figure \ref{fig:number} - \ref{fig:probability}.

In addition, there are many other examples from other industries, such as risk management in financial markets \citep{Yu2018}. In real application, we often take advantage of changes in the topological structure of financial networks to predict oncoming financial crises. It has been found that the above topological structures are usually associated with the co-movements of oil prices. Therefore, the daily spot prices of crude oil are used to measure the risks of financial networks. Thus, this directed network shows the influence of financial activities on the time-dependent cross-correlations of prices at adjacent time points between pairs of regions.  Similar applications include the online monitoring of biodiversity \citep{Fang2013}, periodic orbits embedded in a chaotic attractor \citep{Gao2012},  social networks \citep{Chunaev2020} and other fields.

Univariate control charts, including the Shewhart control chart \citep{Shewhart1925}, the CUSUM control chat \citep{Page1954} and exponentially weighted moving average (EWMA) control chart \citep{Roberts1959}, were designed to signal when an out-of-control (OC) condition came up. With the increasing complexity of production processes, they have been extended to monitor multiple characteristics \citep{Vargas2003, Chenouri2009, Yu2019, Wang2021} and even network cases \citep{Umakanth2013}. For directed networks, \cite{Woodall2017} provided an overview of SPC methods developed for social network data,which mainly considered the shifts in the communication levels of a network \citep{Azarnoush2016} and average closeness and betweenness of a network \citep{McCulloh2011}. \cite{Perry2020} was interested in detecting shifts in the hierarchical tendency of binary digraphs. Time-independence was assumed in these papers, which did not usually match reality, and it has been demonstrated that conventional control schemes introduce many false alarms when the independence assumption is violated \citep{Harris1991}. \cite{Chen2021} presented a matrix autoregressive model in a bilinear form to analyze of dynamic networks. It was limited as it was unable to handle a network with a large number of nodes, since the computational memory required easily exceeds the machine's capacity in this problem.

In the literature, there has been some discussion of process monitoring serially correlated data. Residual control charts, that detect one-step forecast errors, were commonly used on this topic (e.g., \cite{Apley1999, Schmid1997, Loredo2002, Li2009}). However, their performance is sensitive to the assumed parametric time series models that usually do not stand up in practice. Some nonparametric algorithms were also proposed to monitor dynamic data with very undemanding requirements, including the dynamic screening system (DySS) method \citep{Qiu2014, Lijun2016} and top-$r$ control chart \citep{Yu2021}. The former makes use of kernel functions, while the latter mainly detects the shifts in the quasi-maximum likelihood estimator (QMLE). Both are computationally slow diagnostic procedures with many tuning parameters and not very robust detection. To overcome this difficulty, the concept of spring length, which was first discussed in \cite{Chatterjee2009}, has been well developed in control algorithms. See, for instance, \cite{Li2020}, \cite{Qiu2020}, \cite{Xue2021}, and etc. Though it has the significant advantage that only correlation between adjacent points is assumed, it can only be used in the univariate case. In conclusion, the scalability of these methods will certainly provide new opportunities for research in model-free monitoring algorithms in network data.

Our paper concentrates on developing a robust and sensitive control scheme to monitor a doubly correlated directed network. The directed network is modeled by a transition matrix and the transition probability matrix is regarded as the detection object. There are two advantages: weakening the serial correlation and standardizing the data. First, since there is only one type of serial correlation in the transition probability matrix, monitoring it can improve the robustness of control schemes by comparison with monitoring the transfer matrix directly. Second, the sum of each row is one and every element value is between zero and one in a transition probability matrix. Then we can apply the multivariate CUSUM method to each row of data and use a weighted sum of CUSUM statistics to create a new control chart, because they have the same measurement units. The total number of transitions serves as a weighting function to display the importance of each point in the network. Significantly, we extend the spring-length-based method to the multivariate case to solve the unstable performance of the algorithm that arises from the serial correlation. Compared with conventional control charts, this approach has weaker assumptions and can be applied to practical situations more easily. In particular, our novel way only assumes that the process observation is associated with nearby points without any parametric time series models. In brief, it is a new insight into constructing a control chart to monitor the above problem.

The remainder of this paper is organized as follows. In the next section, we develop the spring length based method to solve the serially correlated problem in the multivariate case. Section 3 evaluates the performance of the proposed chart by comparison with the advanced CUSUM and EWMA control charts. In this section, a new CUSUM control chart based on the weighted sum of the above uncorrelated statistic is proposed to monitor the transition probability matrix. Further, numerical simulation studies are conducted and performance comparisons are made in Section 3. In Section 4, we introduce a step-by-step illustration of the proposed algorithm and illustrate the superiority of our design using the example of traffic network monitoring. Several remarks conclude the paper and further studies are proposed in Section 5.

\section{Methodologies}
According to \cite{Kilduff2003}, monitoring the relationship between the reciprocal and transitive tendencies in a directed network is a simple technique to assess its shift. To achieve it, the numbers of transitions in the network are represented digitally as a matrix in Section \ref{sub:2.1}. Then a way to decorrelate the vector is proposed and used in each row of a serially correlated matrix in Section \ref{sub:2.2}. Section \ref{sub:2.3} illustrates a new CUSUM statistic using the above de-correlation method in the multivariate case. Finally, we design a weighted CUSUM control algorithm to detect the mean shift of the digraph in Section \ref{sub:2.4}.

\subsection{Digital representation of digraph}\label{sub:2.1}
\begin{figure}
  \centering
  \includegraphics[scale=0.7]{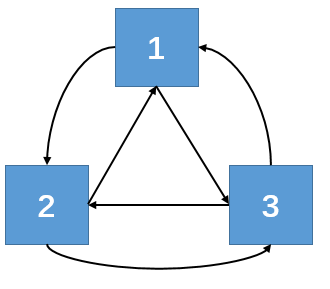}
  \caption{Fully transitive tournament with three vertices}\label{fig:fullytransitation}
\end{figure}

In a complex directed network, there could be a transitive relationship between any two nodes, for example, see Figure \ref{fig:fullytransitation}. $n_{ij}^{(t)}$ denotes the total numbers flowing from node $i$ to node $j$ at time $t$. Then $n_{ij}^{(t)}$ equal to zero indicates no edge flows from node $i$ to node $j$ at time $t$. In this setting, Figure \ref{fig:fullytransitation} suggests a fully transitive tournament with three nodes and it contains all six possible ways to define a transitive flow between the three nodes in \cite{Perry2020}.

Assume that we have $K$ nodes in one digraph, then we can observe a transitive relationship at time $t$ as follows:

\[
\bm{\mathcal{N}}^{(t)}=\left(
             \begin{array}{cccc}
               n_{11}^{(t)} & n_{12}^{(t)} & \cdots & n_{1K}^{(t)} \\
               n_{21}^{(t)} & n_{22}^{(t)} & \cdots & n_{2K}^{(t)} \\
               \cdots & \cdots & \ddots & \cdots \\
               n_{K1}^{(t)} & n_{K2}^{(t)} & \cdots & n_{KK}^{(t)} \\
             \end{array}
           \right).
\]
We call $\bm{\mathcal{N}}^{(t)}$ a transition matrix that can describe a transitive network well.

As described above, $\bm{\mathcal{N}}^{(t)}$ usually is doubly corrected, especially in a complex transport network. In particular, the total numbers that flow out of the node and the transition probability are both serially correlated. Thus a direct analysis on $\bm{\mathcal{N}}^{(t)}$ may not perform robustly. Further, the scale of the $n_{ij}^{(t)}$s ($i=1,2,\cdots, K$, $j=1,2,\cdots, K$) variously reflects the importance of each
node. For instance, the value of $n_{ij}^{(t)}$ is very large when the nodes $i$ and $j$ are active in the network and is small otherwise. To facilitate the following modeling, the transition probability matrix is applied as our measure of transitive relationship here:
\[
\bm{\mathcal{P}}^{(t)}=\left(
             \begin{array}{cccc}
               p_{11}^{(t)} & p_{12}^{(t)} & \cdots & p_{1K}^{(t)} \\
               p_{21}^{(t)} & p_{22}^{(t)} & \cdots & p_{2K}^{(t)} \\
               \cdots & \cdots & \ddots & \cdots \\
               p_{K1}^{(t)} & p_{K2}^{(t)} & \cdots & p_{KK}^{(t)} \\
             \end{array}
           \right),
\]
where $p_{ij}^{(t)}=\frac{n_{ij}^{(t)}}{n_{i}^{(t)}}$ and $n_{i}^{(t)}=\sum\limits_{j=1}^{K}n_{ij}^{(t)}$. Then we have $\sum\limits_{j=1}^{K}p_{ij}^{(t)}=1$ ($i=1,2,\cdots,K$). Thus, the $i$-th row data of $\bm{\mathcal{P}}^{(t)}$, denoted as $\bm{\mathcal{P}}_{i}^{(t)}=(p_{i1}^{(t)} , p_{i2}^{(t)},\cdots , p_{iK}^{(t)})^{T}$, is standardized.

\subsection{De-correlation approach in the multivariate variables}\label{sub:2.2}
Recall that $\bm{\mu}_{i}$ and $\Sigma_{i} $ are the mean vector and the covariance matrix of the $i$-th row data of the above transition probability matrix respectively, where any shift in $\bm{\mu}_{i}$ shows the change in the transfer relationship in the network. For example, the shift in the station-to-station relationship in the metro network prompts operators to make a schedule adjustment in time. Then the problem is equivalent to hypothesis testing:

\begin{equation}
  H_{0}: \bm{\mu}_{i}=\bm{\mu}_{0,i}~~\text{v.s.}~~H_{1}: \bm{\mu}_{i}=\bm{\mu}_{1,i},
\end{equation}
where $\bm{\mu}_{0,i}$ and $\bm{\mu}_{1,i}$ are the in-control mean vector and the expected out-of-control mean vector, respectively, and $\bm{\mu}_{0,i}\neq \bm{\mu}_{1,i}$. Then we use the conventional multivariate CUSUM control chart based on the independence and normality assumptions to handle it as follows:
\begin{equation}\label{eq:ccsum}
\begin{aligned}
C_{i,t}^{+}&=\max(0, C_{i,t-1}^{+}+ e_{i,t} - k_{i})\\
C_{i,t}^{-}&=\min(0, C_{i,t-1}^{-}+ e_{i,t} + k_{i})
\end{aligned},
\end{equation}
where $e_{i,t}=(\bm{\mathcal{P}}_{i}^{(t)}-\bm{\mu}_{0,i})^{T}(\Sigma_{i})^{-1}(\bm{\mu}_{1,i}-\bm{\mu}_{0,i})$, $k_{i}=\frac{1}{2}(\bm{\mu}_{1,i}-\bm{\mu}_{0,i})^{T}(\Sigma_{i})^{-1}(\bm{\mu}_{1,i}-\bm{\mu}_{0,i})$ and $C_{i,0}^{+}=C_{i,0}^{+}=0$.

As described above, the serial correlation of the $e_{i,t}$ necessitates a decorrelation process before a control chart is used. As in \cite{Qiu2020}, we can assume that the observations are covariance stationary and only two points with an interval less than $B$ are serially correlated, where $B>0$ is an integer. More specifically, the autocorrelation coefficient $\bm{\gamma}_{i}(q)= cov(\bm{\mathcal{P}}_{i}^{(t)}, \bm{\mathcal{P}}_{i}^{(t+q)} )$ is a proper measure of the correlation. In practice, $\bm{\gamma}_{i}(q)$ usually decreases with interval $q$ because the data are time-sensitive and observations closer to the current point have greater impact. Thus there exists a proper value of $B$ such that $\gamma_{i}(q)=0$ when $q>B$. Further, $\bm{\gamma}_{i}(q)$ can be estimated by
\begin{equation}
 \hat{\gamma}_{i}(q)=\frac{1}{m-q}\sum\limits_{t=-m+1}^{-q}(\bm{\mathcal{P}}_{i}^{(t)} - \bm{\mu}_{0,i})^{T}(\bm{\mathcal{P}}_{i}^{(t+q)} - \bm{\mu}_{0,i}),
\end{equation}
where $\bm{\mathcal{P}}_{i}^{(-m+1)}, \bm{\mathcal{P}}_{i}^{(-m+2)},\cdots, \bm{\mathcal{P}}_{i}^{(0)}$ are IC data and $m\gg B$.

According to the Cholesky decomposition of the covariance matrices, we propose the $t$-th asymptotically uncorrelated and standardized statistic as
\begin{equation}\label{eq.decorr}
 \tilde{e}_{i,t}=(\bm{\mathcal{P}}_{i}^{(t)}-\bm{\mu}_{0,i}-\hat{\bm{\sigma}}_{i}^{T}\hat{\Gamma}_{i}^{-1}\bm{e}_{i}^{*})^{T}\widetilde{\Sigma}_{i}^{-1}(\bm{\mu}_{1,i}-\bm{\mu}_{0,i}),
\end{equation}
where
\begin{itemize}
  \item \[
   \tilde{\Gamma}_{i}=\left(
                              \begin{array}{ccc}
                                \hat{\gamma}_{i}(0) &  \cdots & \hat{\gamma}_{i}(B) \\
                                \hat{\gamma}_{i}(1) &  \cdots & \hat{\gamma}_{i}(B-1) \\
                                \cdots &  \ddots & \cdots \\
                                \hat{\gamma}_{i}(B)  & \cdots & \hat{\gamma}_{i}(0) \\
                              \end{array}
                            \right)=\left(
                                      \begin{array}{cc}
                                        \hat{\Gamma}_{i} & \hat{\bm{\sigma}}_{i} \\
                                        \hat{\bm{\sigma}}_{i}^{T} & \hat{\gamma}_{i}(0) \\
                                      \end{array}
                                    \right),
    \]
  \item $\bm{e}_{i}^{*}=(\bm{\mathcal{P}}_{i}^{(t-B)}-\bm{\mu}_{0,i}, \cdots, \bm{\mathcal{P}}_{i}^{(t-1)}-\bm{\mu}_{0,i})$,
  \item $\widetilde{\Sigma}_{i}=(1-\hat{\bm{\sigma}}_{i}^{T}\hat{\Gamma}_{i}^{-1}\hat{\bm{\sigma}}_{i})\Sigma_{i}$.
\end{itemize}
The estimation of parameter $B$ will be discussed in the next subsection.

\subsection{The new CUSUM statistic for serially correlated vectors}\label{sub:2.3}
In this subsection, we first introduce the spring-length-based method to estimate the parameter $B$ and then give an improved CUSUM statistic to monitor the row data of the transition probability matrix.

As \cite{Qiu2020} suggested, the spring length based method is an appropriate way to estimate the parameter $B$. The spring length is defined as the time between the current time and the previous time when a CUSUM applied to the original process observations is reset to zero \citep{Chatterjee2009}. It indicates that past observations before the spring length of the current time can be ignored since no more useful information would be provided by them. It encourages us that the parameter $B$ is reset to zero when the CUSUM statistic is equal to zero. In detail, the value of $B$ can be updated by adding one to the original value when the lag between the present and the past is within the spring length, zero otherwise. We notice that the maximum value of $B$, denoted as $B_{max}$, needs to be pre-specified. On the one hand, though the value of $B$ should be chosen to be as large as possible, the estimation of $\gamma(B)$ is more and more unreliable as $B$ increases, especially during the initial period of process monitoring. On the other hand, storing too many past observations is a big challenge for a machine. Of course, the proper value of $B_{max}$ is often application-specific. For example, urban metro transportation systems only preserves the last two hours as the expert advises. Above all, we raise an advanced CUSUM statistic to describe the shift in a serially correlated vector:

\begin{enumerate}
  \item In the case when $t = 1$, the conventional CUSUM statistic is used as
\begin{equation}
\begin{aligned}
\tilde{C}_{i,1}^{+}&=\max\left(0, (\bm{\mathcal{P}}_{i}^{(t)}-\bm{\mu}_{0,i})^{T}(\Sigma_{i})^{-1}(\bm{\mu}_{1,i}-\bm{\mu}_{0,i}) - \frac{1}{2}(\bm{\mu}_{1,i}-\bm{\mu}_{0,i})^{T}(\Sigma_{i})^{-1}(\bm{\mu}_{1,i}-\bm{\mu}_{0,i})\right),\\
\tilde{C}_{i,1}^{-}&=\min\left(0, (\bm{\mathcal{P}}_{i}^{(t)}-\bm{\mu}_{0,i})^{T}(\Sigma_{i})^{-1}(\bm{\mu}_{1,i}-\bm{\mu}_{0,i}) + \frac{1}{2}(\bm{\mu}_{1,i}-\bm{\mu}_{0,i})^{T}(\Sigma_{i})^{-1}(\bm{\mu}_{1,i}-\bm{\mu}_{0,i})\right).
\end{aligned}
\end{equation}
 And two-sided control statistic is defined as $\tilde{C}_{i,1}=\max\left(\tilde{C}_{i,1}^{+}, -\tilde{C}_{i,1}^{-}\right)$. Also define the order of correlation at time $t=1$ as $B_{i,1}=1$ if $C_{i,1}>0$ and 0 otherwise.
  \item In the case when $t \geq 2$, we consider the following two cases:
  \begin{itemize}
    \item If $B_{i,t-1} = 0$, calculate $C_{i,t}$ and $B_{i,t}$ in the same way as that in the case when $t = 1$ discussed above.
    \item If $B_{i,t-1} > 0$, the de-correlation approach in Equation (\ref{eq.decorr}) is used here. Let
    \[
     \tilde{\Gamma}_{i,t}=\left(
                              \begin{array}{ccc}
                                \hat{\gamma}_{i}(0) &  \cdots & \hat{\gamma}_{i}(B_{i,t-1}) \\
                                \hat{\gamma}_{i}(1) &  \cdots & \hat{\gamma}_{i}(B_{i,t-1}-1) \\
                                \cdots &  \ddots & \cdots \\
                                \hat{\gamma}_{i}(B_{i,t-1})  & \cdots & \hat{\gamma}_{i}(0) \\
                              \end{array}
                            \right)=\left(
                                      \begin{array}{cc}
                                        \hat{\Gamma}_{i,t-1} & \hat{\bm{\sigma}}_{i,t-1} \\
                                        \hat{\bm{\sigma}}_{i,t-1}^{T} & \hat{\gamma}_{i}(0) \\
                                      \end{array}
                                    \right).
    \]
Then the new CUSUM statistic are defined as
\begin{equation}\label{eq:decusum}
\begin{aligned}
\tilde{C}_{i,t}^{+}&=\max\left(0, \tilde{C}_{i,t-1}^{+} + \tilde{e}_{i,t} -k_{i}\right),\\
\tilde{C}_{i,t}^{-}&=\min\left(0, \tilde{C}_{i,t-1}^{-} + \tilde{e}_{i,t} +k_{i}\right),
\end{aligned}
\end{equation}
where $ \tilde{e}_{i,t}=(\bm{\mathcal{P}}_{i}^{(t)}-\bm{\mu}_{0,i}-\hat{\bm{\sigma}}_{i,t-1}^{T}\hat{\Gamma}_{i,t-1}^{-1}\bm{e}_{i,t-1}^{*})^{T}\tilde{\Sigma}_{i,t}^{-1}(\bm{\mu}_{1,i}-\bm{\mu}_{0,i})
$, $\bm{e}_{i,t-1}^{*}=(\bm{\mathcal{P}}_{i}^{(t-B_{i,t-1})}-\bm{\mu}_{0,i}, \cdots, \bm{\mathcal{P}}_{i}^{(t-1)}-\bm{\mu}_{0,i})$, and $\tilde{\Sigma}_{i,t}=(1-\hat{\bm{\sigma}}_{i,t-1}^{T}\hat{\Gamma}_{i,t-1}^{-1}\hat{\bm{\sigma}}_{i,t-1})\Sigma_{i}$.

Similarly, $\tilde{C}_{i,t}=\max\left(\tilde{C}_{i,t}^{+}, -\tilde{C}_{i,t}^{-}\right)$ and $B_{i,t}=\begin{cases} \max(B_{max}, B_{i,t-1}+1), & \tilde{C}_{i,t}>0, \\ 0, & \tilde{C}_{i,t}=0.\end{cases}$.
  \end{itemize}
\end{enumerate}

\subsection{The weighted CUSUM control algorithm for whole network detection}\label{sub:2.4}
Here, we design an adaptive weighted CUSUM control algorithm to achieve detection of the whole transition probability matrix.

Though we have achieved row data detection, we cannot simply apply other conventional control charts (e.g., top-$r$ control scheme \citep{{Mei2010}}) to monitor the network because of the endogenous structure of the network. This is one benefit of the transition probability matrix detection, in which the sum of each row is one, in this setting. We can regard each row of data as asymptotically independent observations by excluding the effect of the data size in each node. But obviously the importance of each node is different. We must pay more attention to the shift that happens in a more active node. Thus, we propose an adaptive weighted-sum statistic to detect the whole directed network. At the time $t$, the weighting function and the adaptive weighted-sum statistic are defined as
\begin{equation}
  w_{i}^{(t)}=\frac{n_{i}^{(t)}}{\sum\limits_{i=1}^{K}n_{i}^{(t)}},
\end{equation}

and

\begin{equation}\label{eq:weightedsum}
  \tilde{C}_{t}=\sum\limits_{i=1}^{K}w_{i}^{(t)}\tilde{C}_{i,t}.
\end{equation}

The CUSUM chart defined in Equations (\ref{eq:weightedsum}) gives a signal of mean shift when
\begin{equation}
  \tilde{C}_{t}>h,
\end{equation}
where $h>0$ is the control limit.

In a word, the adaptive weighted CUSUM control algorithm is shown in Algorithm \ref{algo:weightedsum}.

\begin{algorithm}[]
	\caption{The weighted CUSUM control algorithm for serially correlated directed network detection}\label{algo:weightedsum}
		\textbf{Input:} \begin{itemize}
                    \item IC data $\bm{\mathcal{P}}^{(-m+1)},\bm{\mathcal{P}}^{(-m+2)},\cdots, \bm{\mathcal{P}}^{(0)}$;
                    \item IC and excepted OC mean vector of each row data $\bm{\mu}_{0,i}$ and $\bm{\mu}_{1,i}$ ($i=1,2,\cdots,K$) respectively;
                    \item The covariance matrix of each row data $\Sigma_{i}$ ($i=1,2,\cdots,K$);
                    \item The control limit $h$;
                    \item $\tilde{C}_{0}=0$ and $B_{0}=0$.
                  \end{itemize}
		\begin{algorithmic}[1]
        \For {$i=1,2\cdots, K$}
                \State {Compute} $k_{i}=\frac{1}{2}(\bm{\mu}_{1,i}-\bm{\mu}_{0,i})^{T}(\bm{\Sigma}_{i})^{-1}(\bm{\mu}_{1,i}-\bm{\mu}_{0,i})$;
        \EndFor

        \For{ $t=1,2,\cdots$ }
        \State {Observe} $\bm{\mathcal{N}}^{(t)}$;
        \State {Compute} $\bm{\mathcal{P}}^{(t)}$;
        \If {$B_{t-1}=0$}
            \For {$i=1,2\cdots$}
                \State {Compute} $\tilde{e}_{i,t}=(\bm{\mathcal{P}}_{i}^{(t)}-\bm{\mu}_{0,i})^{T}(\bm{\Sigma}_{i})^{-1}(\bm{\mu}_{1,i}-\bm{\mu}_{0,i})$;
            \EndFor
        \Else
             \For {$i=1,2\cdots$}
                \State {Compute} \[
                                     \tilde{\Gamma}_{i,t}=\left(
                                                              \begin{array}{ccc}
                                                                \hat{\gamma}_{i}(0) &  \cdots & \hat{\gamma}_{i}(B_{t-1}) \\
                                                                \hat{\gamma}_{i}(1) &  \cdots & \hat{\gamma}_{i}(B_{t-1}-1) \\
                                                                \cdots &  \ddots & \cdots \\
                                                                \hat{\gamma}_{i}(B_{t-1})  & \cdots & \hat{\gamma}_{i}(0) \\
                                                              \end{array}
                                                            \right)=\left(
                                                                      \begin{array}{cc}
                                                                        \hat{\Gamma}_{i,t-1} & \hat{\bm{\sigma}}_{i,t-1} \\
                                                                        \hat{\bm{\sigma}}_{i,t-1}^{T} & \hat{\gamma}_{i}(0) \\
                                                                      \end{array}
                                                                    \right);
                                    \]
                \State {Compute} $\tilde{\Sigma}_{i,t}=(1-\hat{\bm{\sigma}}_{i,t-1}^{T}\hat{\Gamma}_{i,t-1}^{-1}\hat{\bm{\sigma}}_{i,t-1})\Sigma_{i}$;
                \State {Compute} $ \tilde{e}_{i,t}=(\bm{\mathcal{P}}_{i}^{(t)}-\bm{\mu}_{0,i}-\hat{\bm{\sigma}}_{i,t-1}^{T}\hat{\Gamma}_{i,t-1}^{-1}\bm{e}_{i,t-1}^{*})^{T}\tilde{\Sigma}_{i,t}^{-1}(\bm{\mu}_{1,i}-\bm{\mu}_{0,i})$, where $\bm{e}_{i,t-1}^{*}=(\bm{\mathcal{P}}_{i}^{(t-B_{t-1})}-\bm{\mu}_{0,i}, \cdots, \bm{\mathcal{P}}_{i}^{(t-1)}-\bm{\mu}_{0,i})$ ;
            \EndFor
        \EndIf

        \State {Compute} $\tilde{C}_{i,t}^{+}=\max(0, \tilde{C}_{i,t}^{+}+ \tilde{e}_{i,t}-k_{i})$, $\tilde{C}_{i,t}^{-}=\min(0, \tilde{C}_{i,t}^{-}+ \tilde{e}_{i,t}+k_{i})$ and $\tilde{C_{i,t}}=\max(\tilde{C}_{i,t}^{+}, -\tilde{C}_{i,t}^{-})$;

         \State {Compute} $ \tilde{C}_{t}=\sum\limits_{i=1}^{K}w_{i}^{(t)}\max(\tilde{C}_{i,1}^{+}, -\tilde{C}_{i,1}^{-})$, where $w_{i}^{(t)}=\frac{n_{i}^{(t)}}{\sum\limits_{i=1}^{K}n_{i}^{(t)}}$;
                    \State {Compute} $B_{t}=\begin{cases} \max(B_{max}, B_{t-1}+1 ), & \tilde{C}_{t}>0, \\ 0, & \tilde{C}_{t}=0.\end{cases}$;
        \EndFor
	\end{algorithmic}
	\textbf{Output:} An early warning when $\tilde{C}_{t}>h$. \\
\end{algorithm}

\section{Simulation Performance}
In this section, the numerical performance of the proposed methodology is demonstrated, including the cases with singly and doubly serial correlation, respectively, and the case of correlated row data.

The average and standard deviation of run length are commonly used to evaluate control charts' performance. Concretely, we prefer a long IC average run length ($ARL_{0}$) and a short OC average run length ($ARL_{1}$), while a short standard deviation of run length (SDRL) is desired. The following results are calculated from 10000 replicated simulations.

Besides the proposed control scheme in Algorithm \ref{algo:weightedsum}, denoted as WSCUSUM, the following four types of control charts are considered for comparison:
\begin{enumerate}
  \item The conventional top-$1$ CUSUM chart proposed by \cite{Mei2010}, denoted as TCUSUM, has the charting statistic defined in Equation (\ref{eq:ccsum}). The chart gives a signal when the statistic $\max\limits_{i=1,2,\cdots,K}\max(C_{i,t}^{+}, -C_{i,t}^{-})$ exceeds the control limit.
  \item The de-correlated conventional top-$1$ CUSUM control chart, denoted as DTCUSUM, has the charting statistic defined in Equation (\ref{eq:decusum}). The chart gives a signal when the statistic $\max\limits_{i=1,2,\cdots,K}\max(\tilde{C}_{i,t}^{+}, -\tilde{C}_{i,t}^{-})$ exceeds the control limit.
  \item Like the above de-correlated conventional top-$1$ CUSUM control chart, the control chart, denoted as DTCUSUM-n, is designed to detect $\bm{\mathcal{N}}^{(t)}$ instead of $\bm{\mathcal{P}}^{(t)}$.
  \item The EWMA control chart with applications to social network monitoring proposed by \cite{Perry2020}, denoted as NEWMA, has the charting statistic as
  \begin{equation}
  \begin{aligned}
    G_{i,0}&=0,\\
    G_{i,t}&=(1-\lambda)G_{i,t-1}+\lambda U_{i,t},
  \end{aligned}
  \end{equation}
  where $U_{i,t}=\left(\sum\limits_{j=1}^{K}\frac{p_{ij}^{(t)}}{\mu_{0,ij}\times n_{i}^{(t)}}-K\right)/\sigma^{*}_{i}$, where $\mu_{0,ij}$ is the $j$-th element of $\bm{\mu}_{0,i}$, and $\sigma^{*2}_{i}=\frac{1}{n_{i}^{(t)}}\left(\sum\limits_{j=1}^{K}\frac{1-\mu_{0,ij}}{\mu_{0,ij}}-K(K-1)\right)$.  The upper and lower control limits  are given by $UCL_{t}/LCL_{t}=\pm L\sqrt{\frac{\lambda}{2-\lambda}[1-(1-\lambda)^{2t}]}$, where $L$ is a parameter that can be estimated from the simulation studies in a given $ARL_{0}$ setting. The chart gives a signal when the any statistic $G_{i,t}$ is larger than $UCL_{t}$ or less than $LCL_{t}$. The weighting parameter $\lambda$ is is fixed at 0.1.
\end{enumerate}

Without loss of generality, the IC mean matrix of $\bm{\mathcal{P}}^{(t)}$ is $\left(
                                                                      \begin{array}{cc}
                                                                        0.45 & 0.55 \\
                                                                        0.95 & 0.05 \\
                                                                      \end{array}
                                                                    \right)$,
that demonstrates the two cases of uniform and large transition probability between two nodes, and expected OC mean matrix of $\bm{\mathcal{P}}^{(t)}$ is $\left(
                                                                      \begin{array}{cc}
                                                                        0.5 & 0.5 \\
                                                                        0.5 & 0.5 \\
                                                                      \end{array}
                                                                    \right)$.
According to the suggestion of \cite{Li2020}, the IC dataset of size $m=1000$ is obtained to estimate the IC mean vector, covariance matrix  and correlation coefficient. In the above settings, we design three scenarios, that is $(n_{1}^{(0)},n_{2}^{(0)})=(100,100)$, $(n_{1}^{(0)},n_{2}^{(0)})=(100,500)$, and $(n_{1}^{(0)},n_{2}^{(0)})=(100,30)$, with $ARL_{0}\approx 200$. These cases are studied in the various AR(1) models and we choose $B_{max}=4$ in this paper.

\subsection{The performance in a singly and doubly serial correlation case}\label{sub:3.1}
In this subsection, denote the true mean matrix of $\bm{\mathcal{P}}^{(t)}$ as $\bm{\mu}=\left(
                                                                      \begin{array}{cc}
                                                                        \mu_{11} & 1- \mu_{11} \\
                                                                        1-  \mu_{22}  & \mu_{22} \\
                                                                      \end{array}
                                                                    \right)$. We assume the only one of three, that is $\bm{\mathcal{P}}^{(t)}$, $\bm{\mathcal{N}}^{(t)}$ and $(n_{1}^{(t)},n_{2}^{(t)})$, is serially correlated:
\begin{itemize}
  \item Condition I: Observations $\bm{\mathcal{N}}^{(t)}$ ($t=1,2,\cdots$) follow the AR(1) model:
  \begin{equation}
  \begin{aligned}
   \left(
     \begin{array}{c}
       n_{11}^{(t)} \\
       n_{12}^{(t)} \\
     \end{array}
   \right)&=\max\left(\left(
     \begin{array}{c}
       0 \\
       0 \\
     \end{array}
   \right), \left[0.5\left(
     \begin{array}{c}
       n_{11}^{(t-1)} \\
       n_{12}^{(t-1)} \\
     \end{array}
   \right)+0.5\bm{\epsilon}_{1}\right]\right),\\
   \left(
     \begin{array}{c}
       n_{21}^{(t)} \\
       n_{22}^{(t)} \\
     \end{array}
   \right)&=\max\left(\left(
     \begin{array}{c}
       0 \\
       0 \\
     \end{array}
   \right), \left[0.9\left(
     \begin{array}{c}
       n_{21}^{(t-1)} \\
       n_{22}^{(t-1)} \\
     \end{array}
   \right)+0.1\bm{\epsilon}_{2}\right]\right),\\
  \end{aligned}
  \end{equation}
  where $\bm{\epsilon}_{1}^{(t)}$ and $\bm{\epsilon}_{2}^{(t)}$ ($t=1,2,\cdots$) are independent identically distributed (i.i.d.) with the distribution $N\left((\mu_{11}n_{1},n_{1}-\mu_{11}n_{1})^{T},diag(16,16)\right)$ and $N\left((n_{2}-\mu_{22}n_{2},\mu_{22}n_{2})^{T},diag(16,16)\right)$ respectively, and $[a]$ represents the largest integer not exceeding $a$.

  \item Condition II: Transition probability $\bm{\mathcal{P}}^{(t)}$ ($t=1,2,\cdots$) follow the AR(1) model:
  \begin{equation}\label{eq:arl1p}
  \begin{aligned}
   p_{11}^{(t)}&=\min\left(1, \left(0.5p_{11}^{(t-1)}+0.5\dot{\epsilon}_{1}^{(t)}\right) \vee 0\right),\\
   p_{22}^{(t)}&=\min\left(1, \left(0.9p_{22}^{(t-1)}+0.1\dot{\epsilon}_{2}^{(t)}\right) \vee 0\right),\\
  \end{aligned}
  \end{equation}
  where $a \vee b=\begin{cases}a, a\geq b,\\
  b, a<b,
  \end{cases}$, $\dot{\epsilon}_{1}^{(t)}$ and $\dot{\epsilon}_{2}^{(t)}$ ($t=1,2,\cdots$) are i.i.d. with the distribution $N(p_{11}^{(t-1)},0.0001)$ and $N(p_{22}^{(t-1)},0.0001)$ respectively. Then $ p_{12}^{(t)}=1- p_{11}^{(t)}$ and $ p_{21}^{(t)}=1- p_{22}^{(t)}$. And let $n_{ij}^{(t)}\sim multinomial(n_{i}^{(t)}; p_{ij})$ ($i,j=1,2$).

  \item Condition III: Total number of transition $(n_{1}^{(t)},n_{2}^{(t)})$ follow the AR(1) model:
  \begin{equation}\label{eq:arl1n}
  \begin{aligned}
   n_{1}^{(t)}&=\max(0, [0.5n_{1}^{(t-1)}+0.5\ddot{\epsilon}_{1}^{(t)}]),\\
   n_{2}^{(t)}&=\max(0, [0.9n_{2}^{(t-1)}+0.1\ddot{\epsilon}_{2}^{(t)}]),\\
  \end{aligned}
  \end{equation}
  where $\ddot{\epsilon}_{1}^{(t)}$ and $\ddot{\epsilon}_{2}^{(t)}$ ($t=1,2,\cdots$) are i.i.d. with the distribution $N(n_{1}^{(t-1)},100)$ and $N(n_{2}^{(t-1)},100)$ respectively. Then let $n_{ij}^{(t)}\sim multinomial(n_{i}^{(t)}; p_{ij})$ ($i,j=1,2$).
\end{itemize}

Table \ref{tab:condition1}-\ref{tab:condition3} show that the shift only happens in $(\mu_{11}, 1-\mu_{11})^{T}$ . More extensive results for shifts in other cases will be studied in Section \ref{sub:3.3}.

Firstly, the feasibility of transition probability matrix detection is verified by the simulation results, since SDRLs of the DTCUSUM-n control chart, which monitors the transition matrix only, are usually bigger than the other four control schemes, which monitor the transition probability matrix, whether the shifts occur in the transition number, total number, or transition probability. From Table \ref{tab:condition1}, we can conclude that:
\begin{itemize}
  \item Compared with TCUSUM, DTCUSUM and NEWMA control charts, our proposed method overcomes the non-monotonically decreasing property of the ARLs with increasing shifts, especially in the case of small shifts;
  \item Compared with TCUSUM and NEWMA control charts, our control scheme has smaller SDRLs in small shifts. Moreover, the SDRLs in the TCUSUM and NEWMA are much larger than the ARLs in them, which means that these two control charts are extremely unreliable when they detect small shifts. On the contrary, the WSCUSUM control chart is robust in various cases based on the fact that its SDRLs are always less than the corresponding ARLs;
  \item Our control scheme has the most sensitive performance when monitoring small shifts, especially when the transition numbers are small and $|\mu_{11}-\mu_{0,11}|\leq0.05$, since its ARLs are least;
  \item Increasing transition numbers can correct the slow warning during large shifts.
\end{itemize}
Similar results also can be found in Table \ref{tab:condition2} and Table \ref{tab:condition3}.

When considering the different cases of serial correlation, the robustness of the WSCUSUM control chart is most prominently demonstrated in two ways: (i) The $ARL_{1}$ always becomes smaller with an increasing mean shift
and initial total number of transition; (ii) The SDRLs are always smaller than the corresponding ARLs. Furthermore, compared with the results in Table \ref{tab:condition1} and Table \ref{tab:condition3}, the WSCUSUM control chart performs best in Condition II, where the transition probability matrix is serially correlated, with the smaller ARLs and SDRLs as expected.

Overall, the WCUSUM has unique advantages in terms of robustness. Though it is not the most sensitive in some cases, its performance in ARLs is acceptable.

\begin{sidewaystable}[thp]
\caption{The ARL performance of different control charts in Condition I (the corresponding SDRLs are shown in parentheses).}\label{tab:condition1}\par
\centering
\setlength\tabcolsep{0.3pt}
\renewcommand{\arraystretch}{1.3}
\tiny
\begin{tabular}{r|c|c|c|c|c|c|c|c|c|c|c|c|c|c|c}
  \hline
 \multirow{2}*{$\mu_{11}$} &  \multicolumn{5}{c|} {$(n_{1}^{(0)},n_{2}^{(0)})=(100,100)$}  &  \multicolumn{5}{c|} {$(n_{1}^{(0)},n_{2}^{(0)})=(100,500)$}&  \multicolumn{5}{c} {$(n_{1}^{(0)},n_{2}^{(0)})=(100,30)$}\\\cline{2-16}
& WSCUSUM & TCUSUM & DTCUSUM & DTCUSUM-n & NEWMA & WSCUSUM & TCUSUM & DTCUSUM & DTCUSUM-n & NEWMA & WSCUSUM & TCUSUM & DTCUSUM & DTCUSUM-n & NEWMA \\
  \hline
\multirow{2}*{0.45} & 200.21 & 200.24 & 200.28 & 198.27 & 200.99 & 200.53 & 200.68 & 199.93 & 199.53 & 200.46 & 199.68 & 200.01 & 199.23 & 199.02 & 200.94 \\
& (164.72) & (213.19) & (163.38) & (580.09) & (223.86) & (163.46) & (214.14) & (168.24) & (581.73) & (299.85) & (176.31) & (213.65) & (172.78) & (581.21) & (193.34) \\
 \multirow{2}*{0.449} & 200.01 & 203.51 & 211.43 & 203.67 & 198.48 & 196.00 & 204.03 & 210.50 & 400.74 & 191.62 & 199.63 & 199.10 & 199.34 & 198.53 & 200.38 \\
  & (162.80) & (215.92) & (175.45) & (587.49) & (221.34) & (158.35) & (219.85) & (171.76) & (788.70) & (289.64) & (177.80) & (214.59) & (175.18) & (580.35) & (195.98) \\
 \multirow{2}*{0.448} & 198.82 & 195.62 & 204.04 & 205.17 & 197.24 & 189.44 & 197.73 & 199.63 & 390.54 & 174.31 & 196.81 & 193.88 & 201.04 & 189.77 & 199.62 \\
   & (164.92) & (210.87) & (166.72) & (589.36) & (217.06) & (153.65) & (211.87) & (164.12) & (780.83) & (260.97) & (169.89) & (208.27) & (174.80) & (567.62) & (191.70) \\
\multirow{2}*{0.447} & 162.17 & 179.87 & 160.53 & 209.93 & 197.64 & 168.41 & 181.93 & 168.16 & 399.53 & 145.29 & 194.85 & 177.90 & 201.59 & 187.40 & 198.21 \\
  & (124.99) & (189.16) & (127.94) & (596.12) & (218.12) & (133.01) & (191.80) & (131.20) & (787.79) & (213.13) & (170.45) & (186.62) & (177.30) & (564.51) & (192.99) \\
 \multirow{2}*{0.446} & 144.99 & 166.64 & 144.82 & 203.86 & 197.76 & 152.89 & 169.06 & 152.39 & 392.83 & 132.00 & 184.03 & 165.91 & 198.66 & 200.47 & 201.75 \\
  & (108.37) & (177.94) & (108.58) & (587.76) & (222.07) & (117.45) & (178.82) & (116.35) & (782.75) & (196.78) & (158.91) & (177.60) & (173.65) & (583.12) & (195.60) \\
  \multirow{2}*{0.445} & 117.87 & 147.79 & 116.68 & 198.21 & 195.36 & 128.30 & 148.91 & 128.34 & 400.66 & 108.62 & 171.68 & 147.96 & 198.44 & 195.86 & 200.90 \\
   & (81.59) & (159.71) & (82.46) & (580.11) & (216.90) & (93.01) & (162.52) & (92.33) & (788.99) & (159.10) & (144.23) & (160.05) & (180.07) & (577.09) & (195.79) \\
\multirow{2}*{0.44} & 55.52 & 69.39 & 54.53 & 180.68 & 200.11 & 61.81 & 70.77 & 61.53 & 350.45 & 40.36 & 118.25 & 69.14 & 198.08 & 180.77 & 201.17 \\
  & (28.85) & (75.40) & (28.08) & (558.09) & (221.71) & (33.59) & (77.04) & (33.88) & (749.71) & (54.30) & (76.55) & (75.90) & (173.92) & (558.07) & (196.15) \\
 \multirow{2}*{0.435} & 34.71 & 33.69 & 34.08 & 140.12 & 201.48 & 37.95 & 33.72 & 37.74 & 316.17 & 19.55 & 81.89 & 33.45 & 173.56 & 151.06 & 200.89 \\
  & (14.28) & (35.92) & (14.27) & (496.73) & (226.42) & (17.19) & (36.23) & (17.20) & (720.83) & (23.22) & (39.01) & (35.58) & (120.65) & (514.96) & (195.72) \\
  \multirow{2}*{0.43}& 25.22 & 18.08 & 24.44 & 106.10 & 185.51 & 26.65 & 18.13 & 26.58 & 242.82 & 11.32 & 65.53 & 17.79 & 142.12 & 109.99 & 200.21 \\
   & (8.96) & (18.81) & (8.71) & (436.14) & (205.48) & (10.73) & (18.77) & (10.73) & (645.29) & (11.60) & (26.36) & (18.59) & (77.52) & (443.83) & (197.52) \\
  \multirow{2}*{0.425} & 19.84 & 10.72 & 19.33 & 77.00 & 137.39 & 20.17 & 10.74 & 20.21 & 193.06 & 7.81 & 54.82 & 10.78 & 117.82 & 83.10 & 202.68 \\
  & (6.09) & (10.58) & (5.87) & (373.98) & (143.56) & (7.50) & (10.73) & (7.46) & (583.75) & (7.10) & (20.07) & (10.75) & (54.91) & (388.45) & (194.66) \\
 \multirow{2}*{0.42} & 16.60 & 6.98 & 16.17 & 47.51 & 74.70 & 16.19 & 6.94 & 16.20 & 122.62 & 6.01 & 48.01 & 6.96 & 100.90 & 57.93 & 200.33 \\
  & (4.53) & (6.62) & (4.35) & (293.70) & (70.84) & (5.50) & (6.58) & (5.43) & (472.91) & (5.01) & (16.34) & (6.57) & (41.23) & (325.55) & (193.59) \\
  \multirow{2}*{0.415} & 14.16 & 5.03 & 13.86 & 32.82 & 37.89 & 13.43 & 4.97 & 13.41 & 95.20 & 4.81 & 42.46 & 4.92 & 88.66 & 42.89 & 184.60 \\
   & (3.43) & (4.46) & (3.19) & (243.72) & (30.12) & (4.06) & (4.41) & (4.01) & (419.47) & (3.68) & (13.97) & (4.31) & (33.23) & (280.46) & (179.19) \\
  \multirow{2}*{0.41}& 12.66 & 3.69 & 12.43 & 20.27 & 23.53 & 11.56 & 3.72 & 11.54 & 64.12 & 3.93 & 38.20 & 3.75 & 78.46 & 27.76 & 125.16 \\
  & (2.69) & (3.08) & (2.57) & (189.83) & (15.20) & (3.19) & (3.06) & (3.22) & (345.88) & (2.80) & (12.48) & (3.10) & (26.86) & (224.71) & (111.21) \\
 \multirow{2}*{0.405} & 11.52 & 3.02 & 11.26 & 11.12 & 16.58 & 10.22 & 3.05 & 10.21 & 38.92 & 3.32 & 34.80 & 3.00 & 70.25 & 17.19 & 65.66 \\
   & (2.14) & (2.33) & (1.99) & (136.75) & (9.38) & (2.68) & (2.41) & (2.65) & (269.39) & (2.19) & (11.31) & (2.32) & (23.31) & (174.77) & (49.15) \\
  \multirow{2}*{0.4} & 10.65 & 2.47 & 10.41 & 4.85 & 12.41 & 9.22 & 2.46 & 9.14 & 21.23 & 2.90 & 31.37 & 2.47 & 63.75 & 9.50 & 38.79 \\
   & (1.83) & (1.78) & (1.64) & (82.38) & (6.29) & (2.44) & (1.74) & (2.46) & (196.97) & (1.81) & (9.61) & (1.80) & (19.52) & (126.22) & (22.80) \\
  \multirow{2}*{0.35} & 7.08 & 1.08 & 6.97 & 1.00 & 9.96 & 4.30 & 1.08 & 4.32 & 1.00 & 1.23 & 17.50 & 1.08 & 33.70 & 1.00 & 6.04 \\
 & (0.61) & (0.29) & (0.54) & (0.02) & (4.59) & (0.96) & (0.29) & (0.99) & (0.03) & (0.46) & (3.95) & (0.29) & (6.73) & (0.02) & (1.57) \\
 \multirow{2}*{0.3} & 5.98 & 1.00 & 5.96 & 1.00 & 2.89 & 4.00 & 1.00 & 4.00 & 1.00 & 1.00 & 13.52 & 1.00 & 24.26 & 1.00 & 2.94 \\
  & (0.31) & (0.03) & (0.25) & (0.00) & (0.97) & (0.00) & (0.02) & (0.03) & (0.00) & (0.06) & (2.66) & (0.02) & (3.74) & (0.00) & (0.71) \\
 \multirow{2}*{0.25} & 5.46 & 1.00 & 5.32 & 1.00 & 1.53 & 4.00 & 1.00 & 4.00 & 1.00 & 1.00 & 11.39 & 1.00 & 19.29 & 1.00 & 1.92 \\
   & (0.50) & (0.00) & (0.47) & (0.00) & (0.55) & (0.00) & (0.00) & (0.00) & (0.00) & (0.00) & (2.02) & (0.00) & (2.52) & (0.00) & (0.40) \\
  \multirow{2}*{0.2} & 5.02 & 1.00 & 5.01 & 1.00 & 1.03 & 4.00 & 1.00 & 4.00 & 1.00 & 1.00 & 10.03 & 1.00 & 16.31 & 1.00 & 1.24 \\
   & (0.13) & (0.00) & (0.08) & (0.00) & (0.18) & (0.00) & (0.00) & (0.00) & (0.00) & (0.00) & (1.65) & (0.00) & (1.97) & (0.00) & (0.43) \\
  \multirow{2}*{0.15} & 5.00 & 1.00 & 5.00 & 1.00 & 1.00 & 4.00 & 1.00 & 4.00 & 1.00 & 1.00 & 9.13 & 1.00 & 14.28 & 1.00 & 1.00 \\
   & (0.00) & (0.00) & (0.00) & (0.00) & (0.00) & (0.00) & (0.00) & (0.00) & (0.00) & (0.00) & (1.43) & (0.00) & (1.65) & (0.00) & (0.06) \\
  \multirow{2}*{0.1}& 5.00 & 1.00 & 5.00 & 1.00 & 1.00 & 4.00 & 1.00 & 4.00 & 1.00 & 1.00 & 8.42 & 1.00 & 12.81 & 1.00 & 1.00 \\
   & (0.00) & (0.00) & (0.00) & (0.00) & (0.00) & (0.00) & (0.00) & (0.00) & (0.00) & (0.00) & (1.24) & (0.00) & (1.42) & (0.00) & (0.00) \\
  \multirow{2}*{0.05} & 5.00 & 1.00 & 5.00 & 1.00 & 1.00 & 4.00 & 1.00 & 3.99 & 1.00 & 1.00 & 7.93 & 1.00 & 11.73 & 1.00 & 1.00 \\
  & (0.00) & (0.00) & (0.00) & (0.00) & (0.00) & (0.05) & (0.00) & (0.11) & (0.00) & (0.00) & (1.13) & (0.00) & (1.28) & (0.00) & (0.00) \\
  \hline
\end{tabular}
\end{sidewaystable}

\begin{sidewaystable}[thp]
\caption{The ARL performance of different control charts in Condition II (the corresponding SDRLs are shown in parentheses).}\label{tab:condition2}\par
\centering
\setlength\tabcolsep{0.3pt}
\renewcommand{\arraystretch}{1.3}
\tiny
\begin{tabular}{r|c|c|c|c|c|c|c|c|c|c|c|c|c|c|c}
  \hline
 \multirow{2}*{$\mu_{11}$} &  \multicolumn{5}{c|} {$(n_{1}^{(0)},n_{2}^{(0)})=(100,100)$}  &  \multicolumn{5}{c|} {$(n_{1}^{(0)},n_{2}^{(0)})=(100,500)$}&  \multicolumn{5}{c} {$(n_{1}^{(0)},n_{2}^{(0)})=(100,30)$}\\\cline{2-16}
& WSCUSUM & TCUSUM & DTCUSUM & DTCUSUM-n & NEWMA & WSCUSUM & TCUSUM & DTCUSUM & DTCUSUM-n & NEWMA & WSCUSUM & TCUSUM & DTCUSUM & DTCUSUM-n & NEWMA \\
  \hline
\multirow{2}*{0.45}  & 200.83 & 199.49 & 199.21 & 200.09 & 200.11 & 199.27 & 199.37 & 199.82 & 200.52 & 200.94 & 200.41 & 200.47 & 200.66 & 200.01 & 200.17 \\
 & (177.84) & (192.32) & (565.91) & (168.29) & (206.89) & (169.25) & (191.53) & (168.54) & (247.76) & (205.86) & (170.75) & (194.65) & (204.26) & (195.32) & (206.43) \\
\multirow{2}*{0.449} & 188.96 & 196.70 & 179.31 & 191.87 & 198.37 & 192.03 & 243.12 & 192.74 & 195.92 & 199.37 & 189.62 & 197.69 & 190.27 & 204.22 & 195.30 \\
  & (168.69) & (190.93) & (531.33) & (159.34) & (198.38) & (161.05) & (233.08) & (161.37) & (237.55) & (199.72) & (155.81) & (192.44) & (190.06) & (194.64) & (198.51) \\
 \multirow{2}*{0.448} & 186.51 & 191.40 & 170.13 & 179.90 & 196.85 & 180.10 & 233.34 & 180.56 & 186.06 & 196.56 & 178.51 & 193.31 & 176.42 & 200.74 & 193.57 \\
  & (163.81) & (183.57) & (503.02) & (150.13) & (199.87) & (151.58) & (226.18) & (152.36) & (227.08) & (198.86) & (148.18) & (185.81) & (177.48) & (193.88) & (197.67) \\
 \multirow{2}*{0.447}  & 184.75 & 180.79 & 135.36 & 165.93 & 187.57 & 164.40 & 221.15 & 164.94 & 170.96 & 188.32 & 163.52 & 182.26 & 160.26 & 200.64 & 186.37 \\
  & (165.17) & (175.96) & (431.95) & (140.04) & (191.99) & (140.15) & (213.21) & (141.18) & (211.30) & (189.48) & (135.95) & (178.83) & (158.88) & (194.84) & (191.17) \\
  \multirow{2}*{0.446} & 179.92 & 169.53 & 106.21 & 146.00 & 173.53 & 143.89 & 205.68 & 144.07 & 155.40 & 172.81 & 143.86 & 169.88 & 140.29 & 197.74 & 169.40 \\
  & (161.79) & (162.32) & (359.55) & (119.17) & (175.59) & (117.10) & (202.09) & (117.32) & (192.35) & (175.84) & (113.32) & (161.93) & (135.47) & (190.38) & (177.85) \\
  \multirow{2}*{0.445} & 164.38 & 157.14 & 77.33 & 129.25 & 159.74 & 127.84 & 194.62 & 127.85 & 135.83 & 161.29 & 127.77 & 158.31 & 124.57 & 196.99 & 157.74 \\
  & (146.74) & (151.91) & (265.48) & (100.65) & (163.29) & (101.88) & (192.23) & (100.98) & (162.52) & (162.36) & (100.22) & (153.24) & (119.78) & (183.57) & (161.69) \\
   \multirow{2}*{0.44} & 92.78 & 101.29 & 26.38 & 73.03 & 98.84 & 71.01 & 117.07 & 70.90 & 73.01 & 99.79 & 72.28 & 101.52 & 69.99 & 158.42 & 98.06 \\
 & (74.77) & (95.06) & (73.78) & (48.28) & (96.78) & (47.29) & (113.00) & (47.37) & (85.31) & (97.27) & (46.46) & (94.86) & (54.99) & (120.98) & (96.36) \\
   \multirow{2}*{0.435} & 54.91 & 62.76 & 15.64 & 48.40 & 58.59 & 46.46 & 70.38 & 46.45 & 41.84 & 58.81 & 49.16 & 62.60 & 46.93 & 117.39 & 58.34 \\
& (37.66) & (57.10) & (36.85) & (26.14) & (55.34) & (26.06) & (64.10) & (25.95) & (43.67) & (54.53) & (26.30) & (57.14) & (31.77) & (70.08) & (55.57) \\
  \multirow{2}*{0.43} & 36.69 & 39.76 & 12.19 & 35.88 & 37.14 & 33.99 & 43.45 & 34.05 & 28.21 & 37.12 & 36.89 & 39.65 & 35.24 & 90.13 & 36.35 \\
   & (21.91) & (33.70) & (25.37) & (16.36) & (31.79) & (16.25) & (37.90) & (16.28) & (26.75) & (31.32) & (16.86) & (33.60) & (21.05) & (43.71) & (31.82) \\
   \multirow{2}*{0.425} & 27.58 & 27.31 & 9.68 & 28.73 & 24.94 & 26.90 & 28.99 & 26.95 & 20.58 & 25.01 & 29.69 & 27.37 & 28.70 & 72.50 & 24.84 \\
   & (13.79) & (21.72) & (18.32) & (11.29) & (20.43) & (11.30) & (23.23) & (11.32) & (17.70) & (20.39) & (11.77) & (21.83) & (15.15) & (30.93) & (20.41) \\
   \multirow{2}*{0.42} & 21.99 & 19.79 & 8.28 & 23.82 & 18.35 & 22.17 & 21.03 & 22.16 & 15.93 & 18.28 & 24.77 & 19.57 & 24.20 & 60.97 & 18.29 \\
   & (9.86) & (14.96) & (14.36) & (8.22) & (14.11) & (8.23) & (15.67) & (8.25) & (12.38) & (14.06) & (8.55) & (14.77) & (11.73) & (22.63) & (14.43) \\
  \multirow{2}*{0.415} & 18.43 & 15.30 & 7.42 & 20.71 & 13.93 & 19.04 & 16.09 & 19.04 & 13.03 & 14.05 & 21.72 & 15.19 & 21.17 & 52.58 & 13.79 \\
   & (7.35) & (10.46) & (12.05) & (6.36) & (10.14) & (6.52) & (10.99) & (6.51) & (9.49) & (10.18) & (6.83) & (10.33) & (9.47) & (17.33) & (10.14) \\
   \multirow{2}*{0.41} & 15.93 & 12.36 & 6.62 & 18.39 & 11.25 & 16.68 & 12.77 & 16.64 & 11.01 & 11.24 & 19.23 & 12.37 & 18.72 & 46.29 & 10.91 \\
   & (5.77) & (7.81) & (10.04) & (5.13) & (7.86) & (5.18) & (8.13) & (5.15) & (7.69) & (7.84) & (5.37) & (7.87) & (7.95) & (13.76) & (7.80) \\
   \multirow{2}*{0.405} & 14.02 & 10.13 & 5.94 & 16.68 & 9.04 & 14.94 & 10.49 & 14.95 & 9.46 & 9.12 & 17.43 & 10.01 & 16.97 & 41.34 & 9.00 \\
   & (4.46) & (5.90) & (8.41) & (4.13) & (6.00) & (4.26) & (6.14) & (4.27) & (6.38) & (5.95) & (4.52) & (5.83) & (6.86) & (11.37) & (6.07) \\
   \multirow{2}*{0.4}& 12.65 & 8.53 & 5.66 & 15.18 & 7.61 & 13.52 & 8.88 & 13.53 & 8.32 & 7.59 & 16.03 & 8.55 & 15.56 & 37.29 & 7.52 \\
   & (3.70) & (4.59) & (7.67) & (3.39) & (4.90) & (3.54) & (4.73) & (3.53) & (5.36) & (4.86) & (3.78) & (4.59) & (5.94) & (9.56) & (4.93) \\
  \multirow{2}*{0.35} & 7.11 & 3.43 & 3.41 & 9.33 & 2.48 & 7.59 & 3.55 & 7.61 & 3.91 & 2.50 & 9.81 & 3.41 & 9.57 & 19.90 & 2.47 \\
   & (1.24) & (1.13) & (2.17) & (0.99) & (1.30) & (1.18) & (1.16) & (1.18) & (1.66) & (1.30) & (1.31) & (1.12) & (2.64) & (2.96) & (1.30) \\
  \multirow{2}*{0.3} & 5.05 & 2.23 & 3.02 & 7.59 & 1.42 & 5.78 & 2.31 & 5.79 & 3.09 & 1.41 & 7.94 & 2.23 & 7.76 & 14.30 & 1.41 \\
   & (1.37) & (0.52) & (0.42) & (0.55) & (0.60) & (0.65) & (0.55) & (0.64) & (0.40) & (0.59) & (0.78) & (0.53) & (1.73) & (1.44) & (0.59) \\
   \multirow{2}*{0.25} & 3.24 & 1.80 & 3.00 & 6.94 & 1.07 & 4.95 & 1.87 & 4.95 & 2.89 & 1.07 & 6.97 & 1.80 & 6.90 & 11.55 & 1.06 \\
  & (1.02) & (0.43) & (0.11) & (0.25) & (0.26) & (0.37) & (0.38) & (0.38) & (0.32) & (0.26) & (0.55) & (0.43) & (1.34) & (0.86) & (0.25) \\
  \multirow{2}*{0.2} & 2.35 & 1.34 & 3.00 & 6.09 & 1.00 & 4.27 & 1.44 & 4.26 & 2.33 & 1.00 & 6.59 & 1.34 & 6.50 & 9.92 & 1.00 \\
   & (0.52) & (0.48) & (0.00) & (0.28) & (0.06) & (0.44) & (0.50) & (0.44) & (0.47) & (0.07) & (0.67) & (0.47) & (1.11) & (0.54) & (0.05) \\
   \multirow{2}*{0.15} & 2.02 & 1.04 & 3.00 & 6.00 & 1.00 & 4.00 & 1.06 & 4.00 & 2.01 & 1.00 & 5.92 & 1.04 & 6.03 & 8.92 & 1.00 \\
  & (0.15) & (0.20) & (0.00) & (0.00) & (0.00) & (0.05) & (0.24) & (0.06) & (0.09) & (0.00) & (0.41) & (0.19) & (0.98) & (0.36) & (0.00) \\
   \multirow{2}*{0.1} & 2.00 & 1.00 & 3.00 & 6.00 & 1.00 & 4.00 & 1.00 & 4.00 & 2.00 & 1.00 & 5.90 & 1.00 & 5.69 & 8.00 & 1.00 \\
   & (0.00) & (0.02) & (0.00) & (0.00) & (0.00) & (0.00) & (0.04) & (0.00) & (0.00) & (0.00) & (0.44) & (0.01) & (0.72) & (0.14) & (0.00) \\
   \multirow{2}*{0.05}& 2.00 & 1.00 & 3.00 & 6.00 & 1.00 & 4.00 & 1.00 & 4.00 & 2.00 & 1.00 & 5.87 & 1.00 & 5.68 & 7.38 & 1.00 \\
   & (0.00) & (0.00) & (0.00) & (0.00) & (0.00) & (0.00) & (0.00) & (0.00) & (0.00) & (0.00) & (0.50) & (0.00) & (0.73) & (0.49) & (0.00) \\
   \hline
\end{tabular}
\end{sidewaystable}

\begin{sidewaystable}[thp]
\caption{The ARL performance of different control charts in Condition III (the corresponding SDRLs are shown in parentheses).}\label{tab:condition3}\par
\centering
\setlength\tabcolsep{0.3pt}
\renewcommand{\arraystretch}{1.3}
\tiny
\begin{tabular}{r|c|c|c|c|c|c|c|c|c|c|c|c|c|c|c}
  \hline
 \multirow{2}*{$\mu_{11}$} &  \multicolumn{5}{c|} {$(n_{1}^{(0)},n_{2}^{(0)})=(100,100)$}  &  \multicolumn{5}{c|} {$(n_{1}^{(0)},n_{2}^{(0)})=(100,500)$}&  \multicolumn{5}{c} {$(n_{1}^{(0)},n_{2}^{(0)})=(100,30)$}\\\cline{2-16}
& WSCUSUM & TCUSUM & DTCUSUM & DTCUSUM-n & NEWMA & WSCUSUM & TCUSUM & DTCUSUM & DTCUSUM-n & NEWMA & WSCUSUM & TCUSUM & DTCUSUM & DTCUSUM-n & NEWMA \\
  \hline
\multirow{2}*{0.45} & 199.74 & 199.62 & 200.93 & 199.80 & 200.59 & 200.08 & 199.83 & 200.90 & 199.42 & 199.61 & 199.95 & 200.69 & 200.93 & 197.20 & 200.70 \\
& (56.66) & (350.33) & (545.96) & (598.43) & (203.57) & (211.44) & (371.47) & (224.29) & (597.55) & (206.64) & (56.45) & (361.23) & (521.66) & (594.93) & (206.61) \\
\multirow{2}*{0.449}  & 167.71 & 157.85 & 203.28 & 178.40 & 197.85 & 111.30 & 163.06 & 109.78 & 185.81 & 194.65 & 165.72 & 158.48 & 211.20 & 181.20 & 196.21 \\
& (42.08) & (294.51) & (552.12) & (568.65) & (198.24) & (86.58) & (323.26) & (87.53) & (578.86) & (194.59) & (41.41) & (296.37) & (535.40) & (572.68) & (198.89) \\
\multirow{2}*{0.448} & 145.14 & 121.93 & 195.60 & 156.20 & 176.28 & 71.49 & 139.61 & 70.69 & 157.58 & 171.14 & 141.79 & 120.47 & 190.07 & 167.20 & 173.95 \\
 & (32.88) & (229.00) & (545.44) & (535.11) & (168.19) & (46.62) & (291.69) & (45.33) & (536.95) & (160.55) & (31.35) & (224.90) & (513.55) & (552.09) & (170.41) \\
\multirow{2}*{0.447}  & 121.21 & 87.90 & 180.74 & 129.00 & 143.18 & 46.65 & 96.15 & 46.03 & 154.94 & 138.60 & 117.04 & 87.07 & 177.26 & 130.60 & 139.76 \\
& (24.92) & (181.44) & (528.30) & (489.53) & (129.68) & (25.75) & (214.79) & (24.75) & (532.85) & (126.64) & (23.21) & (178.39) & (501.38) & (492.37) & (131.72) \\
\multirow{2}*{0.446}  & 114.27 & 71.41 & 170.05 & 130.80 & 134.08 & 41.11 & 80.22 & 40.71 & 136.53 & 129.67 & 109.81 & 69.36 & 182.63 & 123.40 & 134.03 \\
 & (23.27) & (131.51) & (516.69) & (492.72) & (120.40) & (21.26) & (171.16) & (20.68) & (502.45) & (116.17) & (20.81) & (121.92) & (509.73) & (479.42) & (124.36) \\
\multirow{2}*{0.445} & 103.72 & 55.49 & 186.70 & 100.80 & 116.78 & 33.47 & 61.39 & 33.52 & 130.73 & 114.01 & 98.79 & 54.54 & 166.77 & 105.60 & 116.78 \\
& (19.92) & (91.94) & (542.43) & (435.50) & (103.60) & (15.73) & (126.92) & (15.46) & (492.34) & (100.39) & (17.55) & (88.34) & (488.00) & (445.28) & (105.30) \\
\multirow{2}*{0.44}  & 72.23 & 17.70 & 146.76 & 48.60 & 65.71 & 17.72 & 18.99 & 17.78 & 81.06 & 65.18 & 66.24 & 17.62 & 116.96 & 51.40 & 64.22 \\
& (13.38) & (17.88) & (490.49) & (304.87) & (55.57) & (5.99) & (19.93) & (5.99) & (391.93) & (55.04) & (8.96) & (17.70) & (416.90) & (313.48) & (54.93) \\
\multirow{2}*{0.435}  & 56.77 & 7.43 & 82.79 & 22.00 & 43.15 & 12.36 & 8.54 & 12.31 & 48.84 & 42.43 & 50.56 & 7.46 & 7.13 & 22.00 & 43.37 \\
& (11.17) & (6.68) & (309.36) & (203.87) & (35.99) & (3.25) & (7.75) & (3.23) & (305.49) & (34.46) & (6.08) & (6.84) & (12.77) & (203.87) & (36.50) \\
\multirow{2}*{0.43}  & 47.06 & 3.85 & 35.99 & 8.20 & 30.75 & 9.70 & 4.57 & 9.71 & 20.82 & 29.95 & 41.31 & 3.81 & 4.10 & 8.80 & 30.36 \\
 & (10.35) & (3.12) & (136.10) & (119.79) & (25.08) & (2.01) & (3.79) & (2.03) & (198.02) & (24.56) & (4.54) & (3.13) & (4.88) & (124.66) & (24.97) \\
\multirow{2}*{0.425}  & 40.12 & 2.39 & 18.27 & 3.80 & 23.43 & 8.16 & 2.84 & 8.16 & 12.21 & 23.50 & 35.04 & 2.37 & 3.08 & 4.40 & 23.17 \\
   & (9.46) & (1.69) & (76.03) & (74.78) & (19.23) & (1.40) & (2.06) & (1.39) & (149.25) & (19.50) & (3.65) & (1.65) & (2.86) & (82.40) & (19.22) \\
\multirow{2}*{0.42} & 34.28 & 1.72 & 9.87 & 1.20 & 18.98 & 7.20 & 2.05 & 7.19 & 6.60 & 18.72 & 30.48 & 1.70 & 2.52 & 1.60 & 18.63 \\
  & (8.73) & (1.00) & (45.56) & (20.00) & (15.65) & (1.04) & (1.27) & (1.03) & (105.69) & (15.97) & (2.88) & (1.00) & (1.75) & (34.64) & (15.45) \\
\multirow{2}*{0.415}  & 29.06 & 1.35 & 5.78 & 1.40 & 15.67 & 6.52 & 1.59 & 6.52 & 4.00 & 15.34 & 27.11 & 1.36 & 2.24 & 1.40 & 15.57 \\
 & (7.49) & (0.63) & (28.39) & (28.28) & (13.35) & (0.80) & (0.84) & (0.80) & (77.41) & (13.03) & (2.46) & (0.65) & (1.08) & (28.28) & (13.44) \\
\multirow{2}*{0.41} & 24.72 & 1.17 & 4.04 & 1.20 & 13.15 & 6.02 & 1.33 & 6.03 & 2.00 & 12.92 & 24.48 & 1.17 & 2.10 & 1.00 & 13.00 \\
 & (5.38) & (0.42) & (19.05) & (20.00) & (11.33) & (0.66) & (0.60) & (0.66) & (44.71) & (11.17) & (2.11) & (0.42) & (0.65) & (0.00) & (11.40) \\
 \multirow{2}*{0.405}  & 21.65 & 1.07 & 2.66 & 1.00 & 11.13 & 5.68 & 1.18 & 5.67 & 2.00 & 10.98 & 22.33 & 1.07 & 2.04 & 1.00 & 11.11 \\
 & (3.31) & (0.27) & (10.07) & (0.00) & (9.76) & (0.59) & (0.42) & (0.58) & (44.71) & (9.64) & (1.83) & (0.27) & (0.37) & (0.00) & (9.89) \\
\multirow{2}*{0.4}  & 19.60 & 1.03 & 2.25 & 1.00 & 9.74 & 5.37 & 1.08 & 5.37 & 1.20 & 9.58 & 20.56 & 1.03 & 2.01 & 1.00 & 9.74 \\
 & (1.82) & (0.17) & (5.85) & (0.00) & (8.70) & (0.51) & (0.29) & (0.51) & (20.00) & (8.61) & (1.59) & (0.16) & (0.23) & (0.00) & (8.68) \\
\multirow{2}*{0.35}  & 12.14 & 1.00 & 2.00 & 1.00 & 3.68 & 2.29 & 1.00 & 2.28 & 1.00 & 3.59 & 12.53 & 1.00 & 2.00 & 1.00 & 3.65 \\
 & (0.55) & (0.00) & (0.00) & (0.00) & (3.70) & (0.58) & (0.00) & (0.57) & (0.00) & (3.63) & (0.58) & (0.00) & (0.00) & (0.00) & (3.63) \\
\multirow{2}*{0.3} & 9.65 & 1.00 & 2.00 & 1.00 & 2.02 & 2.00 & 1.00 & 2.00 & 1.00 & 2.05 & 9.95 & 1.00 & 2.00 & 1.00 & 2.07 \\
& (0.48) & (0.00) & (0.00) & (0.00) & (2.02) & (0.00) & (0.00) & (0.00) & (0.00) & (2.13) & (0.31) & (0.00) & (0.00) & (0.00) & (2.17) \\
\multirow{2}*{0.25} & 8.14 & 1.00 & 2.00 & 1.00 & 1.47 & 2.00 & 1.00 & 2.00 & 1.00 & 1.46 & 8.72 & 1.00 & 2.00 & 1.00 & 1.47 \\
 & (0.35) & (0.00) & (0.00) & (0.00) & (1.37) & (0.00) & (0.00) & (0.00) & (0.00) & (1.32) & (0.45) & (0.00) & (0.00) & (0.00) & (1.34) \\
\multirow{2}*{0.2}& 7.54 & 1.00 & 2.00 & 1.00 & 1.25 & 2.00 & 1.00 & 2.00 & 1.00 & 1.23 & 7.96 & 1.00 & 2.00 & 1.00 & 1.27 \\
   & (0.50) & (0.00) & (0.00) & (0.00) & (1.02) & (0.00) & (0.00) & (0.00) & (0.00) & (0.93) & (0.19) & (0.00) & (0.00) & (0.00) & (1.07) \\
 \multirow{2}*{0.15} & 7.00 & 1.00 & 2.00 & 1.00 & 1.16 & 2.00 & 1.00 & 2.00 & 1.00 & 1.15 & 7.02 & 1.00 & 2.00 & 1.00 & 1.15 \\
 & (0.05) & (0.00) & (0.00) & (0.00) & (0.77) & (0.00) & (0.00) & (0.00) & (0.00) & (0.74) & (0.14) & (0.00) & (0.00) & (0.00) & (0.79) \\
 \multirow{2}*{0.1}& 6.86 & 1.00 & 2.00 & 1.00 & 1.13 & 2.00 & 1.00 & 2.00 & 1.00 & 1.10 & 7.00 & 1.00 & 2.00 & 1.00 & 1.11 \\
 & (0.35) & (0.00) & (0.00) & (0.00) & (0.67) & (0.00) & (0.00) & (0.00) & (0.00) & (0.59) & (0.04) & (0.00) & (0.00) & (0.00) & (0.62) \\
\multirow{2}*{0.05} & 6.00 & 1.00 & 2.00 & 1.00 & 1.09 & 2.00 & 1.00 & 2.00 & 1.00 & 1.09 & 6.87 & 1.00 & 2.00 & 1.00 & 1.09 \\
 & (0.02) & (0.00) & (0.00) & (0.00) & (0.54) & (0.00) & (0.00) & (0.00) & (0.00) & (0.54) & (0.34) & (0.00) & (0.00) & (0.00) & (0.56) \\
   \hline
\end{tabular}
\end{sidewaystable}

\subsection{The performance in a doubly serial correlation case}\label{sub:3.2}
In this subsection, we study the control schemes' performance in case of doubly serial correlation, that is the total number and probability of transition are both serially correlated. The following model is considered: $\bm{\mathcal{P}}^{(t)}$ ($t=1,2,\cdots$) follows the model in Equation (\ref{eq:arl1p}), while $(n_{1}^{(t)},n_{2}^{(t)})$  ($t=1,2,\cdots$) comes from the model in Equation (\ref{eq:arl1n}). Then let $n_{ij}^{(t)}\sim multinomial(n_{i}^{(t)}; p_{ij})$ ($i,j=1,2$).

\begin{figure}[h!]
  \centering
  \includegraphics[scale=0.4]{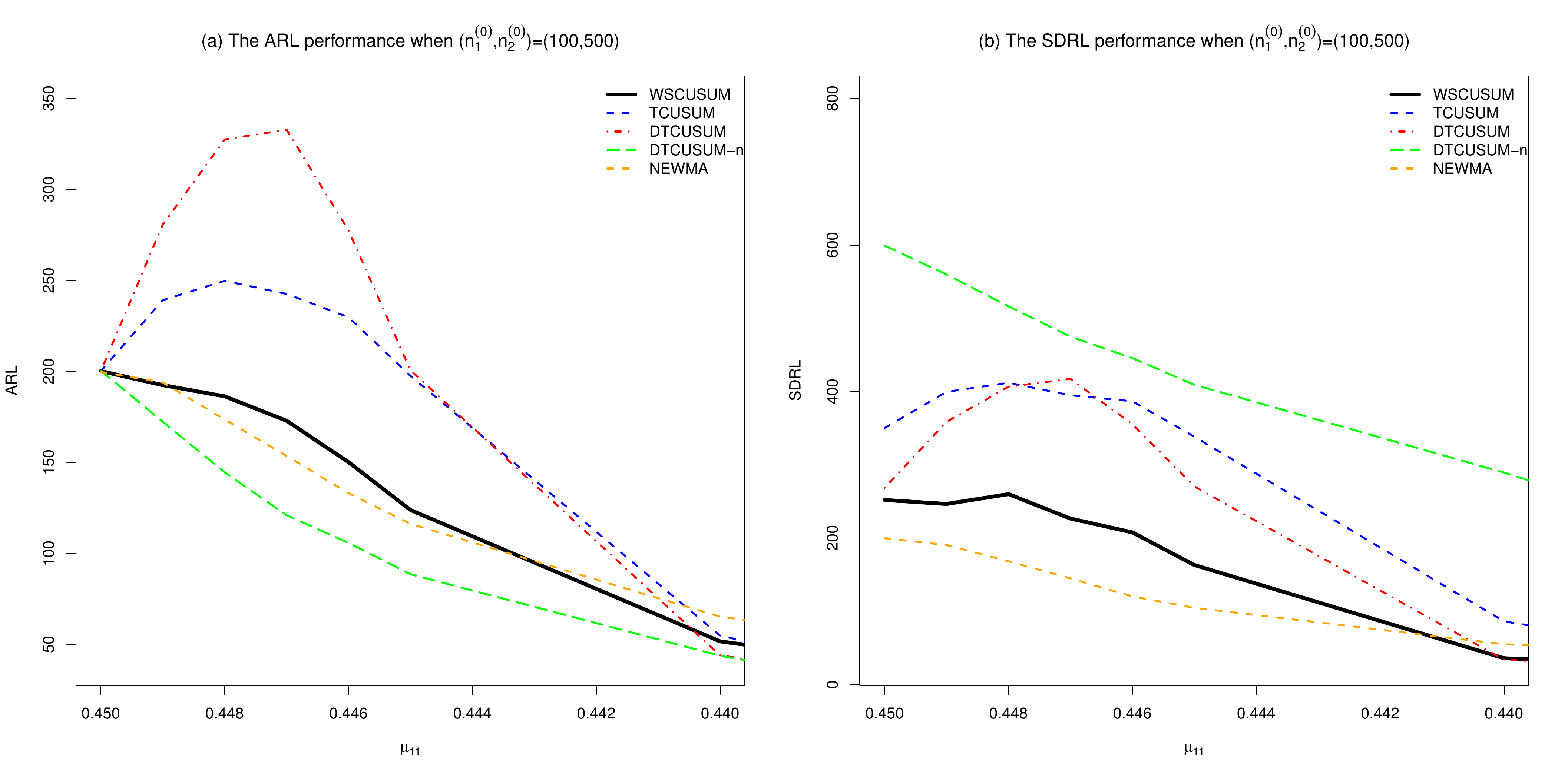}
  \caption{The performance of different control charts when the total number and probability of transition are both serially correlated.}\label{fig:double}
\end{figure}

The values of ARLs and SDRLs when the initial total numbers of transition are equal to $(100,500)$ are visualized in Figures \ref{fig:double}. The very similar results when  $(n_{1}^{(0)}, n_{2}^{(0)})=(100,100)$ and $(n_{1}^{(0)}, n_{2}^{(0)})=(100,30)$ are available upon request. From Figure \ref{fig:double}, we can see that:
\begin{itemize}
  \item The TCUSUM control chart is most sensitive in the case with the smallest ARLs. The WSCUSUM and NEWMA control charts behave similarly for their own specialty. The WSCUSUM performs better than NEWMA when shifts are somewhat larger and vice-versa. The DTCUSUM and DTCUSUM-n control charts are the worst, with the largest and nonmonotonic ARLs.
  \item The WSCUSUM and NEWMA control charts are the most robust of the five control charts since their SDRLs are the smallest. Though the TCUSUM control chart has the smallest ARLs, it is less robust and has the largest SDRLs. We notice that the value of SDRL is almost three times larger than ARL. The big values of SDRLs in the DTCUSUM and DTCUSUM-n control charts are not to be  ignored lightly.
\end{itemize}
To sum up, the WSCUSUM and NEWMA control charts are better when the total number and probability of transition are both serially correlated, with comparatively small ARLs and SDRLs at the same time. The former performs better when the shifts are large, while the latter behaves better when the shifts are small.

\subsection{The performance in the case of correlated row data}\label{sub:3.3}
In this subsection, we discuss a much more complex and realistic case, that is the $\bm{\mathcal{P}}_{1}^{(t)}$ and $\bm{\mathcal{P}}_{2}^{(t)}$ are correlated. To describe this feature, we assume the following model:
\begin{equation}\label{eq:arl1pp}
  \begin{aligned}
   p_{11}^{(t)}&=\min\left(1, \left(0.5p_{11}^{(t-1)}+0.5\underline{\epsilon}_{1}^{(t)}\right) \vee 0\right),\\
   p_{22}^{(t)}&=\min\left(1, \left(0.9p_{11}^{(t-1)}+0.1\underline{\epsilon}_{2}^{(t)}\right) \vee 0\right),\\
  \end{aligned}
  \end{equation}
  where $\underline{\epsilon}_{1}^{(t)}$ and $\underline{\epsilon}_{2}$ ($t=1,2,\cdots$) are i.i.d. with the distribution $N(p_{11}^{(t-1)},0.0001)$. Then $ p_{12}^{(t)}=1- p_{11}^{(t)}$ and $ p_{21}^{(t)}=1- p_{22}^{(t)}$. Similarly, $(n_{1}^{(t)},n_{2}^{(t)})$ ($t=1,2,\cdots$) comes from the model in Equation (\ref{eq:arl1n}) and $n_{ij}^{(t)}\sim multinomial(n_{i}^{(t)}; p_{ij})$ ($i,j=1,2$).

\begin{figure}[h!]
  \centering
  \includegraphics[scale=0.4]{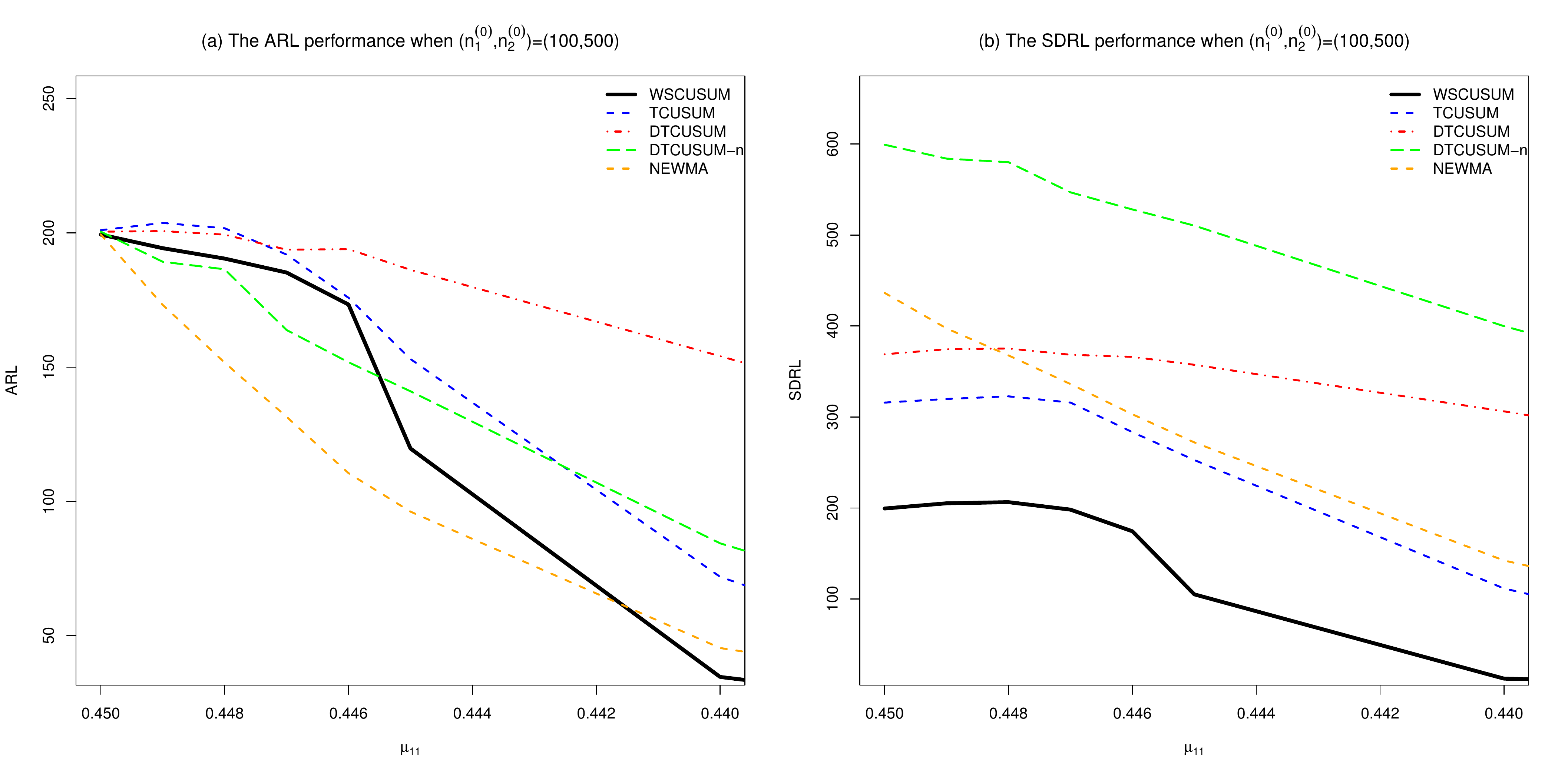}
  \caption{The performance of different control charts when the total number and probability of transition are both serially correlated and row data are correlated.}\label{fig:pcorr}
\end{figure}

Figure \ref{fig:pcorr} displays that our novel approach is the most robust with the smallest SDRLs. The SDRLs of the other four charts are much larger than their ARLs, especially when the shifts are small. Though our method is not the most sensitive when shifts are small, it behaves best when the shifts are somewhat larger and also has an acceptable performance when shifts are small. Finally, compared with the results in Section \ref{sub:3.1} and Section \ref{sub:3.2}, the more complex the network, the more obvious the advantages of the WSCUSUM control chart.

\section{Metro traffic example revisited}
Online traffic monitoring is crucial to allow operating managers to make appropriate evaluations of the current health of the system and plan for the future. In this subsection, the example describing in the Introduction is revisited to demonstrate the application of the proposed control chart.

The data set is drawn from a big data platform for smart transport applications, including the user ID, ticket records, transaction time, in-gate/out-gate station and so on.
 \begin{table}[thp]
\caption{The samples of the traffic data collected from the big data platform.}\label{tab:demo}\par
\centering
\setlength\tabcolsep{0.3pt}
\begin{tabular}{c|c|c|c|c|c|c}
  \hline
Octopus ID &	Transaction Time	&Transaction Type&  In-gate Station	& Out-gate Station	& Train Direction &	Fare\\ \hline
900125532 &	2012/12/31 18:27&ENT	&29	&29	&1	&0\\
900125532	&2012/12/31 18:47&	USE&29&	49&2&10.5\\\hline
900125582	&2012/12/31 18:18&	ENT	&27&	27	&1	&0\\
900125582	&2012/12/31 18:44&	USE&	27&	48&	2&	10.5\\\hline
900125585	&2012/12/31 6:45&	ENT&49&	49&	1&	0\\
900125585	&2012/12/31 7:24&	USE	&	49&	97&	2&	9.7\\\hline
900125600	&2012/12/31 19:07&	ENT&6	&6&	1&	0\\
900125600	&2012/12/31 19:14&	USE	&6	&18&	2	&4.9\\\hline
900125603	&2012/12/31 10:52&	ENT	&	6&	6&1	&0\\
900125603	&2012/12/31 11:02&	USE&	6&	3&2&	4.9\\
   \hline
\end{tabular}
\end{table}
Table \ref{tab:demo} displays a few sample rows of the data collected from the big data platform. Considering the operator's requirements for monitoring, the number of transactions per half hour is used to detect abnormal behavior. We detect a traffic problem on a particular line, that consists of fourteen stations, from 06:00 to 23:30 every day and the data analysis proceeds in several steps as follows:

\begin{itemize}
  \item \textbf{Step 1: Traffic measurement.}

  At the ending of every half hour, the number of transitions from one station to another station is counted and the matrix $\bm{\mathcal{N}}^{(t)}$ is obtained. Data management is then performed and we obtain $\bm{\mathcal{P}}^{(t)}$.

  \item \textbf{Step 2:  Phase I establishment.}

 The company met a big traffic problem in 2019 because of some social events. In this paper, the records from 2017 are chosen to form Phase I (the in-control stage). We use the first eleven months' data to estimate parameters such as mean vector, covariance matrix, correlation coefficient and control limits. Guided by the data provider, the value of $B_{max}$ is set to four here, which means that past observations more than 2 hours before the current time provide little useful information. We assume that the $ARL_{0}=200$ so that a false alarm would be given weekly in the in-control process. The control limits of the five control charts mentioned in Section 3 are shown in Table \ref{tab:real-controllimits}.

 \begin{table}[h!]
\caption{Control limits of real-data example}\label{tab:real-controllimits}\par
\centering
\tabcolsep=3.5pt
\begin{tabular}{c|ccccc}
\hline
Control charts& WSCUSUM & TCUSUM  &  DCUSUM & DCUSUM-n& NEWMA\\\hline
h/L& 638.5 & 62.63 & 18300&0.28&17.26\\
\hline
\end{tabular}
\end{table}

 \item \textbf{Step 3:  Phase I back-testing.}

 Intuitively, the situation of serial correlation raises the false alarm rate of the conventional control charts. To demonstrate the robustness of our proposed WSCUSUM control chart, we test the five control ones in IC data from December 7 to December 13 in 2017. Figure \ref{fig:phasei} shows that the conventional control charts without de-correlation (e.g., TCUSUM and NEWMA ones) both raise an alert, while the WSCUSUM and DCUSUM do not make false alerts. As for the DCUSUM-n control chart, it only gives a signal at the first observation. Thus, the detection of transition number matrix does not work in a complex directed network. In this way, we conclude that our model performs more robustly than the others.
\begin{figure}
  \centering
  \includegraphics[scale=0.5]{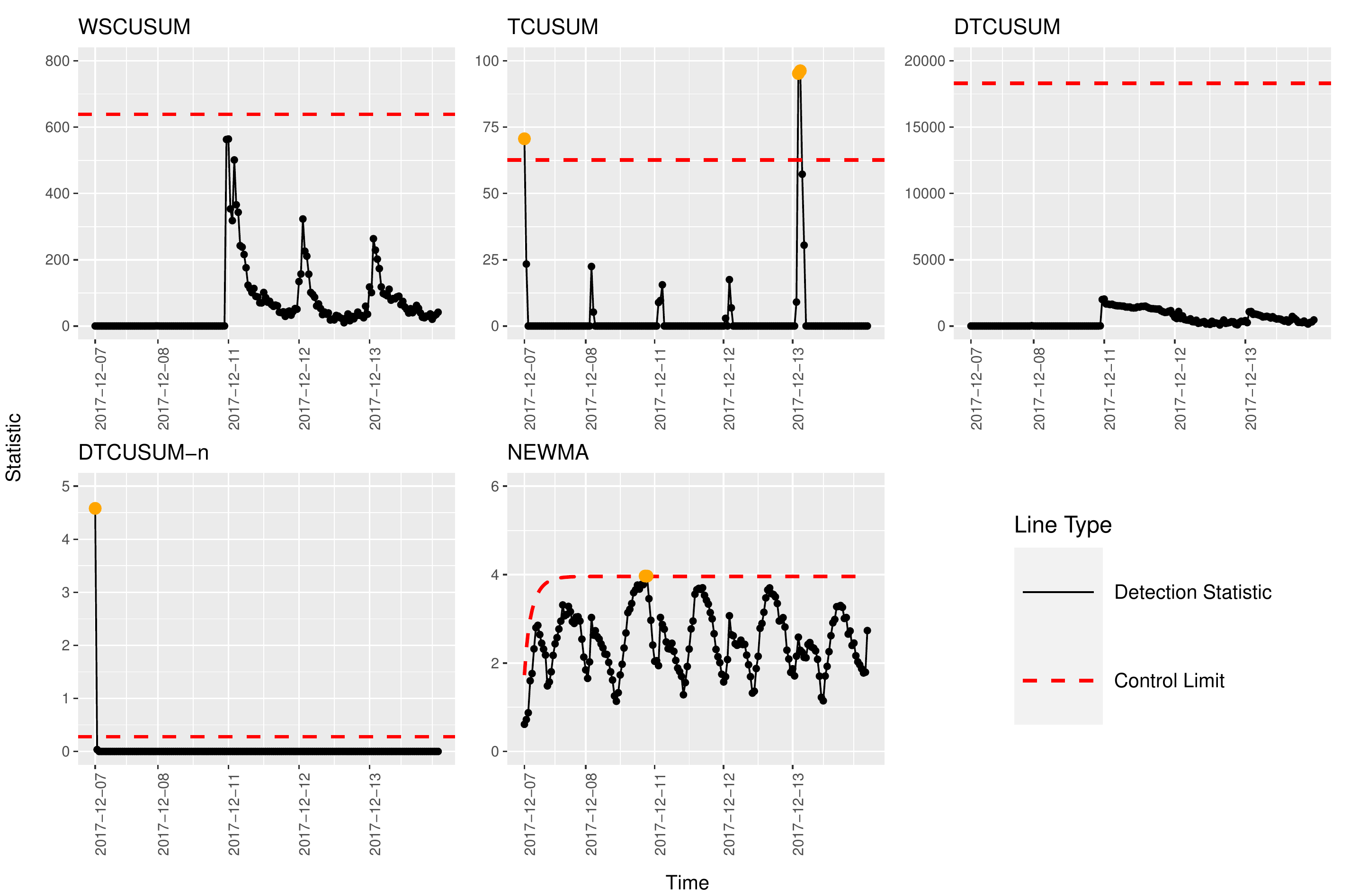}
  \caption{Phase I back-testing performances of five models}\label{fig:phasei}
\end{figure}

\item \textbf{Step 4:  Phase II monitoring.}

In this step, we monitor the OC data from November 11 to November 15 in 2019 to display the sensitivity of our proposed WSCUSUM control chart. As Figure \ref{fig:phaseii} illustrates, the WSCUSUM and TCUSUM control charts signal at the 4th observation. Slightly later, the DCUSUM control charts signal at the 19th observation. As for the DCUSUM-n and NEWMA control chart, the cumulative statistics do not increase with the persistent shifts. Thus, our method is most effective during Phase II.

\begin{figure}
  \centering
  \includegraphics[scale=0.5]{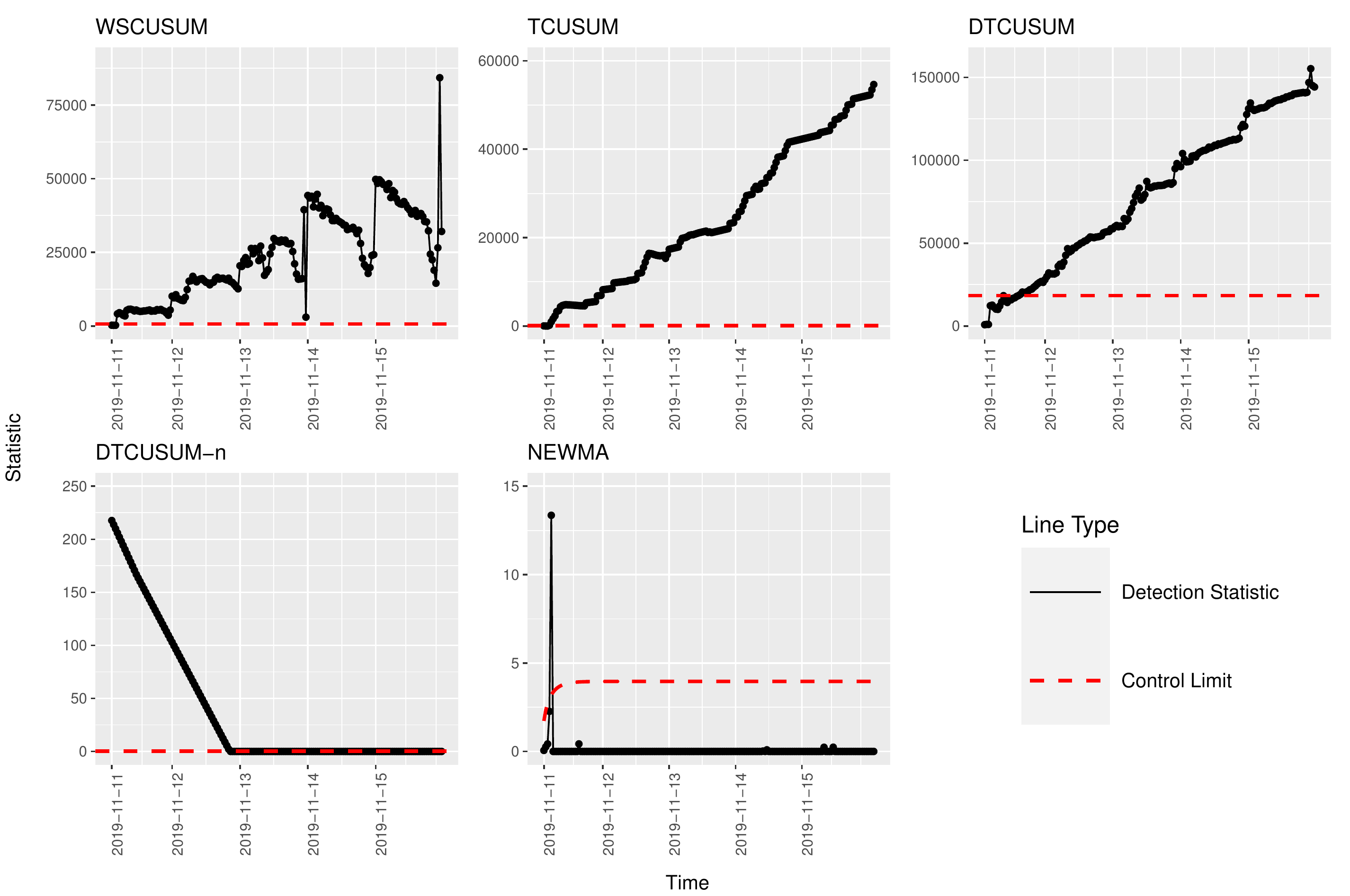}
  \caption{Phase II performance of five models}\label{fig:phaseii}
\end{figure}
\end{itemize}

Above all, the proposed WSCUSUM scheme performs more robustly and sensitively.

\section{Conclusion}
In this paper, prompted by an interesting real metro traffic detection problem, we propose an innovative weighted CUSUM control chart using a spring-length-based decorrelation approach to monitor serially correlated directed networks. The dynamic features are common in real applications, especially when the time difference between the two adjacent points is small. Moreover, in a network, the doubly serial correlation of the transition numbers makes the problem more complex. Thus, we propose to detect shifts in the transition probability matrix rather than the transition matrix to improve the robustness of the algorithm. We extend the spring-length-based decorrelation approach without any parametric assumptions to the multivariate vectors and design independent test statistics. Finally, in consideration of the importance of nodes, a weighted CUSUM control scheme is suggested, in which we regard the total number of transitions at each node as the weighting function. Both simulation results and a  a real-world example of monitoring the mean shifts in metro traffic illustrate that our method is robust and sensitive, and in particular the SDRLs are always small enough. In summary, our charting method is a powerful tool for many process monitoring applications because of its great flexibility.

Finally, we think there are still many new opportunities for research into approaches to robust and sensitive monitoring algorithms in serially correlated networks. On the one hand, a large sample size is required in estimating the $\gamma(B)$ ($B\leq B_{max}$) especially when the value of $B_{max}$ and the number of nodes are large. A de-correlation method in the case where the IC sample size $m$ is small will be developed in further research. On the other hand, though we mainly focus on the case of detecting small shifts, we must admit that our method is not as sensitive as the conventional control charts in monitoring large shifts. An improved method that strikes a balance between robustness and sensitivity in the case of complex serial correlation can be considered.
\normalem
\nocite*{}  

\end{document}